\documentclass[aps,prd,twocolumn,showpacs,amsmath,amssymb]{revtex4-1}
\usepackage{natbib} % 加载 natbib 包
\usepackage{amsmath}
\usepackage{graphicx}
\usepackage{subfigure}
\usepackage{epstopdf}
\usepackage{color}
\usepackage{microtype}
\usepackage{multirow}
\usepackage{setspace}
\usepackage{overpic}
\usepackage{amssymb}
\usepackage{lineno}
\usepackage{bm}
\usepackage{rotating}
\usepackage[colorlinks,urlcolor=blue, citecolor=blue,linkcolor=blue]
{hyperref}
\graphicspath{{./IMG/}}

\let\oldequation\equation
\let\oldendequation\endequation
\renewenvironment{equation}
  {\linenomathNonumbers\oldequation}
  {\oldendequation\endlinenomath}
\newcommand{\BESIIIorcid}[1]{\href{https://orcid.org/#1}{\hspace*{0.1em}\raisebox{-0.45ex}{\includegraphics[width=1em]{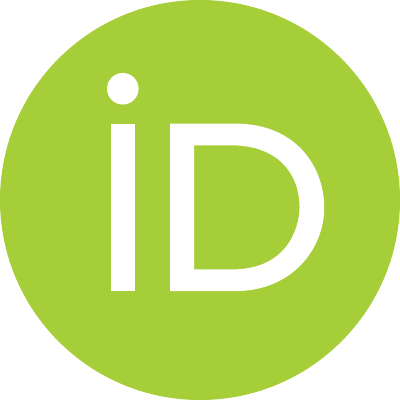}}}}

\begin{document}

\title{\bf \boldmath
Search for $e^{+}e^{-} \to \pi^0 K^+K^-J/\psi$ at $\sqrt{s}= 4.26 \sim$ 4.95 GeV}

\author{
	%% Saved at => 2026-06-05
M.~Ablikim$^{1}$\BESIIIorcid{0000-0002-3935-619X},
M.~N.~Achasov$^{4,c}$\BESIIIorcid{0000-0002-9400-8622},
P.~Adlarson$^{84}$\BESIIIorcid{0000-0001-6280-3851},
X.~C.~Ai$^{90}$\BESIIIorcid{0000-0003-3856-2415},
C.~S.~Akondi$^{32A,32B}$\BESIIIorcid{0000-0001-6303-5217},
R.~Aliberti$^{40}$\BESIIIorcid{0000-0003-3500-4012},
A.~Amoroso$^{83A,83C}$\BESIIIorcid{0000-0002-3095-8610},
Q.~An$^{79,66,\dagger}$,
M.~S.~Anderson$^{40}$\BESIIIorcid{0009-0008-1550-2632},
Y.~Bai$^{64}$\BESIIIorcid{0000-0001-6593-5665}
% ...
\\
\vspace{0.2cm}
(BESIII Collaboration)\\
\vspace{0.2cm} {\it
$^{1}$ Institute of High Energy Physics, Beijing 100049, People's Republic of China\\
% ...
$^{4}$ Budker Institute of Nuclear Physics SB RAS (BINP), Novosibirsk 630090, Russia\\
% ...
$^{32}$ INFN Sezione di Ferrara, (A)INFN Sezione di Ferrara, I-44122, Ferrara, Italy; (B)University of Ferrara, I-44122, Ferrara, Italy\\
% ...
$^{40}$ Johannes Gutenberg University of Mainz, Johann-Joachim-Becher-Weg 45, D-55099 Mainz, Germany\\
% ...
$^{64}$ Southeast University, Nanjing 211100, People's Republic of China\\
% ...
$^{66}$ State Key Laboratory of Particle Detection and Electronics, Beijing 100049, Hefei 230026, People's Republic of China\\
% ...
$^{79}$ University of Science and Technology of China, Hefei 230026, People's Republic of China\\
% ...
$^{83}$ University of Turin and INFN, (A)University of Turin, I-10125, Turin, Italy; (B)University of Eastern Piedmont, I-15121, Alessandria, Italy; (C)INFN, I-10125, Turin, Italy\\
$^{84}$ Uppsala University, Box 516, SE-75120 Uppsala, Sweden\\
% ...
$^{90}$ Zhengzhou University, Zhengzhou 450001, People's Republic of China\\

\vspace{0.2cm}
$^{\dagger}$ Deceased\\
% ...
$^{c}$ Also at the Novosibirsk State University, Novosibirsk, 630090, Russia\\
% ...
}
%% ends here %%

}

\begin{abstract}
Using $e^+e^-$ collision data collected with the BESIII detector at 27 center-of-mass energies between 4.26 and 4.95 GeV corresponding to a total integrated luminosity of $13.0~\text{fb}^{-1}$, we report the first search for the process $e^{+}e^{-} \to \pi^0 K^+K^-J/\psi$. We additionally probe the cross section spectra for potential charmonium-like vector states and invariant mass distributions for intermediate $Z_c/Z_{cs}$ state signals, and find no resonant structures consistent with these exotic candidates. Given the absence of observed resonant signals, upper limits at the 90\% confidence level are determined for the Born cross sections, along with their best-fit values, as well as for the production of the intermediate states.

\end{abstract}

\maketitle

\section*{Introduction}
In recent decades, a series of exotic hadronic states, collectively termed the $XYZ$ states, have been discovered in the charmonium mass region by experiments including Belle, BaBar, BESIII, and CLEO~\cite{Brambilla:2019esw}. Notable examples include the $X(3872)$~\cite{Belle:2003nnu}, $Y(4260)$~\cite{BaBar:2005hhc,CLEO:2006tct,Belle:2007dxy,BaBar:2012vyb}, and $Z_{c}(3900)$~\cite{BESIII:2013ris,Belle:2013yex}.~Their underlying nature remains elusive, prompting a variety of exotic interpretations such as hybrid mesons~\cite{Zhu:2005hp,Close:2005iz,Kou:2005gt,Brambilla:2022hhi,Oncala:2017hop}, tetraquark states~\cite{Maiani:2014aja}, hadronic molecules~\cite{Chen:2017abq,Ding:2008gr,Wang:2013cya,Cleven:2013mka}, hadrocharmonia~\cite{Li:2013ssa,Dubynskiy:2008mq}, or kinematically induced threshold effects~\cite{Chen:2017uof}. Clarifying their properties has spurred extensive experimental and theoretical research in hadron physics~\cite{Wang:2025dur}.

Among these charmonium-like states, the $Y$ states with quantum numbers $J^{PC}=1^{--}$ are of particular interest. States such as the $Y(4260)$, $Y(4360)$~\cite{BaBar:2006ait,Belle:2007umv,BaBar:2012hpr,BESIII:2021njb}, and $Y(4660)$~\cite{Belle:2007umv,BaBar:2012hpr,BESIII:2021njb} exhibit a strong coupling to hidden-charm final states and have been observed prominently in the processes $e^+e^- \to \pi^+ \pi^- J/\psi$ and $e^+e^- \to \pi^+ \pi^- \psi(2S)$. Observations have been extended to the neutral pion channels $e^+e^- \to \pi^0 \pi^0 J/\psi$~\cite{BESIII:2015cld,BESIII:2020oph} and to $e^+e^- \to \pi^+ \pi^- h_{c}$~\cite{CLEO:2011aa,BESIII:2016adj} by the BESIII and CLEO collaborations. Recently, the BESIII experiment reported a three-resonance structure in the same $e^+e^- \to \pi^+ \pi^- h_{c}$ process within the center-of-mass (c.m.)~energy range from 4.009 to 4.950 GeV~\cite{BESIII:2025bce}. Beyond processes involving $\pi\pi$ transitions, final states containing strange mesons have also been studied, including $e^+e^- \to K\bar{K}J/\psi$~\cite{Belle:2016bew,BESIII:2018iop,BESIII:2022joj,BESIII:2023wqy,BESIII:2022kcv}, $e^+e^- \to K^+K^-\psi(2S)$~\cite{BESIII:2024vjf}, and $e^+e^- \to \phi\chi_{cJ}~(J=0,1,2)$~\cite{BESIII:2022wjl,BESIII:2024itg}. In the $e^+e^- \to K^+K^-J/\psi$ process, the BESIII experiment observed a new structure near 4.710 GeV, denoted as $Y(4710)$~\cite{BESIII:2023wqy}. This structure may correspond to the $Y(4660)$ resonance or represent a distinct $Y$ state. Search for $Y$ states have also been conducted in open-charm final states via $e^+e^-$ annihilation. The $Y(4660)$ state decaying to $D_{s}^+D_{s1}(2536)^-$~\cite{Belle:2019qoi} was first observed by the Belle collaboration with a significance of $5.9\sigma$. Three $Y$ states ($Y(4260), Y(4500)$ and $Y(4660)$) were observed for the first time in the process $e^+e^- \to D^{*0} D^{*-} \pi^+$~\cite{BESIII:2023cmv} by the BESIII collaboration. A potential structure around 4.79 GeV was reported in the $D_{s}^{*+}D_{s}^{*-}$ final state~\cite{BESIII:2023wsc}. Whether this structure is related to the $Y(4710)$ remains unclear, and additional data in this energy region are needed to clarify its nature. Studies of $Y$-state decays into hidden-charm final states containing $J/\psi$, $\psi(2S)$, $\chi_{cJ}$, $h_{c}$, or other charmonium states provide crucial insights into their internal structure.

Beyond these neutral vector states, several charged charmonium-like resonances have also been discovered. The $Z_c(3900)^{\pm}$ was first observed in the $\pi^\pm J/\psi$ invariant mass spectrum of the process $e^+e^- \to \pi^+\pi^- J/\psi$ by the BESIII experiment in 2013~\cite{BESIII:2013ris}, with subsequent confirmation by Belle~\cite{Belle:2013yex}, CLEO~\cite{Xiao:2013iha}, and the D0 collaboration~\cite{D0:2019zpb}. The neutral partner, $Z_{c}(3900)^0$, was later observed by BESIII in the decay to $\pi^0 J/\psi$~\cite{BESIII:2015cld,BESIII:2020oph}. Furthermore, the BESIII experiment reported the first observation of the $Z_{c}$ strangeness partner with strange quark content, $Z_{cs}(3985)$, in the process $e^+e^- \to K^+(D^{-}_{s}D^{*0}+D_{s}^{*-}D^{0})$~\cite{BESIII:2020qkh,Yang:2020nrt,Meng:2020ihj,Maiani:2021tri,Chen:2022asf}. A neutral structure consistent with~$Z_{cs}(3985)$ was also reported in the $K_{S}^{0}$ recoil mass spectra by the process $e^+e^- \to K_{S}^{0}(D^{+}_{s}D^{*-}+D_{s}^{*+}D^{-})$~\cite{BESIII:2022qzr}. Concurrently, the LHCb experiment announced the observation of two states, $Z_{cs}(4000)$ and $Z_{cs}(4220)$, in the $K^+J/\psi$ invariant mass spectrum from $B^+ \to J/\psi\phi K^+$ decays~\cite{LHCb:2021uow}. Further investigations into potential decay channels to both hidden-charm and open-charm final states are essential to explore the new $Z_{cs}$ states and to better understand the properties of the existing ones.

Using a data sample of $e^+e^-$ collisions at c.m.~energies between 4.40 and 4.95 GeV, corresponding to a total integrated luminosity of 8.9 fb$^{-1}$, the BESIII collaboration recently reported the first observation of the process $e^{+}e^{-}\to \pi^{+}K^{-}K^{0}J/\psi + \text{c.c.}$~with a statistical significance of 9.4$\sigma$~\cite{BESIII:2025fvc}.

In this paper, we present the first measurement of the Born cross section for $e^{+}e^{-} \to \pi^{0}K^{+}K^{-}J/\psi$.~The analysis is based on a data set corresponding to an integrated luminosity of 13.0 fb$^{-1}$ collected with the BESIII detector at c.m.~energies from 4.26 to 4.95 GeV~\cite{BESIII:2022dxl,BESIII:2022ulv}. The $\pi^0$ and $J/\psi$ candidates are reconstructed via their decays to $\gamma\gamma$ and $l^+l^-$ ($l=e,\mu$), respectively.

\section{BESIII detector and Monte Carlo}

The BESIII detector records symmetric $e^+e^-$ collisions provided by the BEPCII storage ring~\cite{Yu:IPAC2016-TUYA01} in the c.m.~energy range from 1.84 to 4.95 GeV, with a peak luminosity of $1.1 \times10^{33}\;\text{cm}^{-2}\text{s}^{-1}$ achieved at $\sqrt{s} = 3.773\;\text{GeV}$. The cylindrical core of the BESIII detector covers 93\% of the full solid angle and consists of a helium-based multilayer drift chamber~(MDC), a plastic scintillator time-of-flight system~(TOF), and a CsI(Tl) electromagnetic calorimeter~(EMC), which are all enclosed in a superconducting solenoidal magnet providing a 1.0~T magnetic field. The solenoid is supported by an octagonal flux-return yoke with resistive plate counter muon identification modules interleaved with steel~(MUC). The charged particle momentum resolution at $1~{\rm GeV}/c$ is $0.5\%$, and the ${\rm d}E/{\rm d}x$ resolution is $6\%$ for electrons from Bhabha scattering. The EMC measures photon energies with a resolution of $2.5\%$ ($5\%$) at $1$~GeV in the barrel (end cap) region. The time resolution in the TOF barrel region is 68~ps, while that in the end cap region is 110~ps. The end cap TOF system was upgraded in 2015 using multi-gap resistive plate chamber technology, providing a time resolution of 60~ps. About 87\% of the data used here benefits from this upgrade.  

Simulated data samples produced with a {\sc geant4}-based~\cite{GEANT4:2002zbu} Monte Carlo (MC) package that includes the geometric description of the BESIII detector and its detector response, are used to determine detection efficiencies, evaluate the initial state radiation (ISR) correction factor, $(1 + \delta)_{\rm ISR}$ and estimate backgrounds. For the signal processes $e^+e^- \to \pi^0K^+K^-J/\psi$, 50000 signal MC events are generated for each c.m.~energy using a phase space (PHSP) model. The simulation incorporates the beam energy spread and ISR in the $e^+e^-$ annihilation, modelled with the generator {\sc kkmc}~\cite{Jadach:1999vf} and {\sc evtgen}~\cite{Ping:2008zz}. A helix parameter correction for charged tracks is applied during kinematic fits to improve consistency between the performance of MC and data. The inclusive MC sample includes the production of open-charm processes, ISR production of vector charmonium(-like) states, and continuum processes, all incorporated via {\sc kkmc}. Inclusive MC samples are generated with integrated luminosities corresponding to ten times that of the actual data at the c.m.~energies of 4.26~GeV, 4.36~GeV, and 4.60~GeV. For the continuous energy regions, samples with forty times and ten times the data luminosity are produced for the 4.60 $\sim$ 4.70~GeV and 4.74  $\sim$ 4.95~GeV intervals, respectively. The decays of generated particles are modeled with the {\sc evtgen} package, employing branching fractions either documented by the Particle Data Group (PDG)~\cite{ParticleDataGroup:2026mpi}, when available, or otherwise estimated with {\sc lundcharm}~\cite{Yang:2014vra}. Final state radiation from charged particles is incorporated using the {\sc photos} package~\cite{Richter-Was:1992hxq}.

\section{Event selection}

In this analysis, a partial-reconstruction strategy is employed to mitigate the low reconstruction efficiency for low-momentum charged kaons. Events are retained in which the $\pi^0$ and $J/\psi$ candidates are fully reconstructed through the $\gamma\gamma$ and $l^+l^-$ decay modes, respectively, while only one or two of the charged kaons in the final state are required to be detected. Charged tracks detected in the MDC are required to be within $|\cos\theta| < 0.93$, where $\theta$ is the polar angle defined with respect to the symmetry axis of the MDC. The distance of closest approach for each track to the interaction point (IP) is required to be less than 10~cm along the beam direction ($|V_z| < 10$~cm) and less than 1~cm in the transverse plane ($|V_{xy}| < 1$~cm). Charged tracks with momentum greater than 1.0~GeV/$c$ are identified as leptons, and all others are treated as kaon candidates. To distinguish between electron and muon candidates, a cut is applied based on the ratio $E/p$, defined as the EMC deposited energy over the candidate's reconstructed momentum. Candidates with $E/p > 0.6$ are identified as electrons, while those with $E/p < 0.6$ are classified as muons. To purify the kaon sample, particle identification (PID) is applied, combining measurements of the specific ionization energy loss ($\mathrm{d}E/\mathrm{d}x$) in the MDC and the flight time in the TOF. A charged track is identified as a kaon if it satisfies the requirement ${\mathcal P}_K > {\mathcal P}_\pi$ and ${\mathcal P}_K > 0$, where ${\mathcal P}_{(K,\pi,...)}$ denotes the PID probability for a given particle hypothesis. Photon candidates are selected from isolated showers in the EMC. The deposited energy must be greater than 25~MeV in the barrel region ($|\cos\theta| < 0.80$) and greater than 50~MeV in the end cap region ($0.86 < |\cos\theta| < 0.92$). To suppress electronic noise and showers unrelated to the event, the difference between the EMC time and the event start time is required to be within [0, 700]~ns. The number of selected photon candidates per event must satisfy $1 < N_\gamma < 7$.

For events with more than two photon candidates, the combination yielding the least $\chi^{2}$ of the kinematic fit is selected. For events in the full reconstruction mode ($e^{+}e^{-} \to \gamma\gamma K^+K^-l^+l^-$), a four-constraint (4C) kinematic fit is applied to enforce four-momentum conservation. The fit quality is required to satisfy $\chi^{2}_{\rm 4C} < 55$. This requirement is optimized by maximizing a Figure-of-Merit (FOM) defined as $S/(\alpha/2 +\sqrt{B})$~\cite{chiopt}, where $\alpha = 3$ is the target significance, and $S$ and $B$ are the expected numbers of signal and background events obtained from signal and inclusive MC samples, respectively. For events in the partial reconstruction mode, a one-constraint (1C) kinematic fit is performed under the hypothesis $e^{+}e^{-} \to \gamma\gamma K^{\pm}_{\rm miss}K^{\mp}l^+l^-$, where $K_{\rm miss}^{\pm}$ denotes a missing charged kaon and the mass recoiling against the $\gamma\gamma K^{\pm}l^+l^-$ system is constrained to the nominal kaon mass~\cite{ParticleDataGroup:2026mpi}. The corresponding requirement on the fit quality is $\chi_{\rm 1C}^{2} < 10$. For the $\mu^+\mu^-$ final state, where background from $\mu/\pi$ misidentification is significant, the muon identification criteria are tightened: the hit depth in the muon counter is required to be greater than 24~cm and 34~cm for the full and partial reconstruction modes, respectively. These values are determined via the FOM optimization method.

\begin{figure*}[!htb]
	\centering
	\includegraphics[angle=0,width=0.42\textwidth]{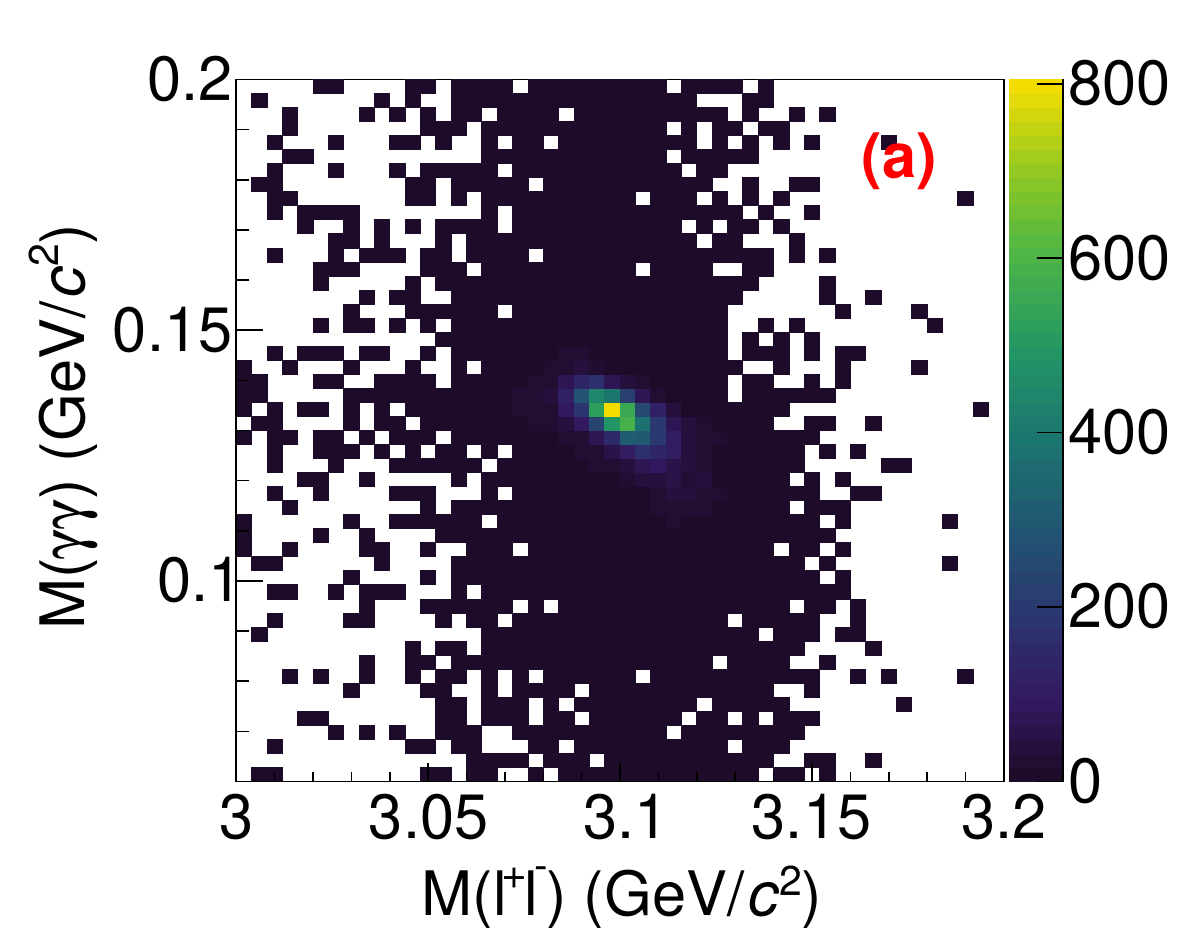}
	\includegraphics[angle=0,width=0.42\textwidth]{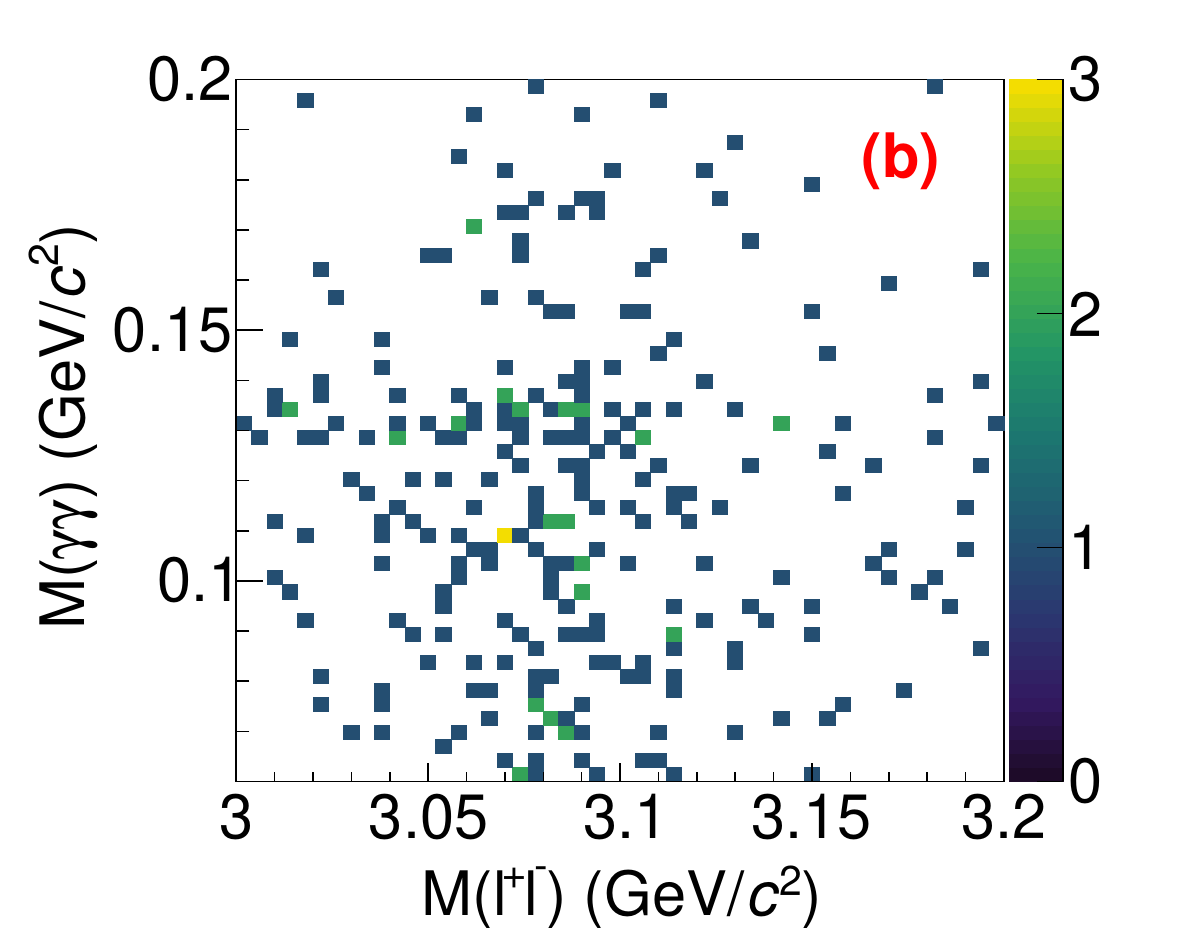}
	\includegraphics[angle=0,width=0.42\textwidth]{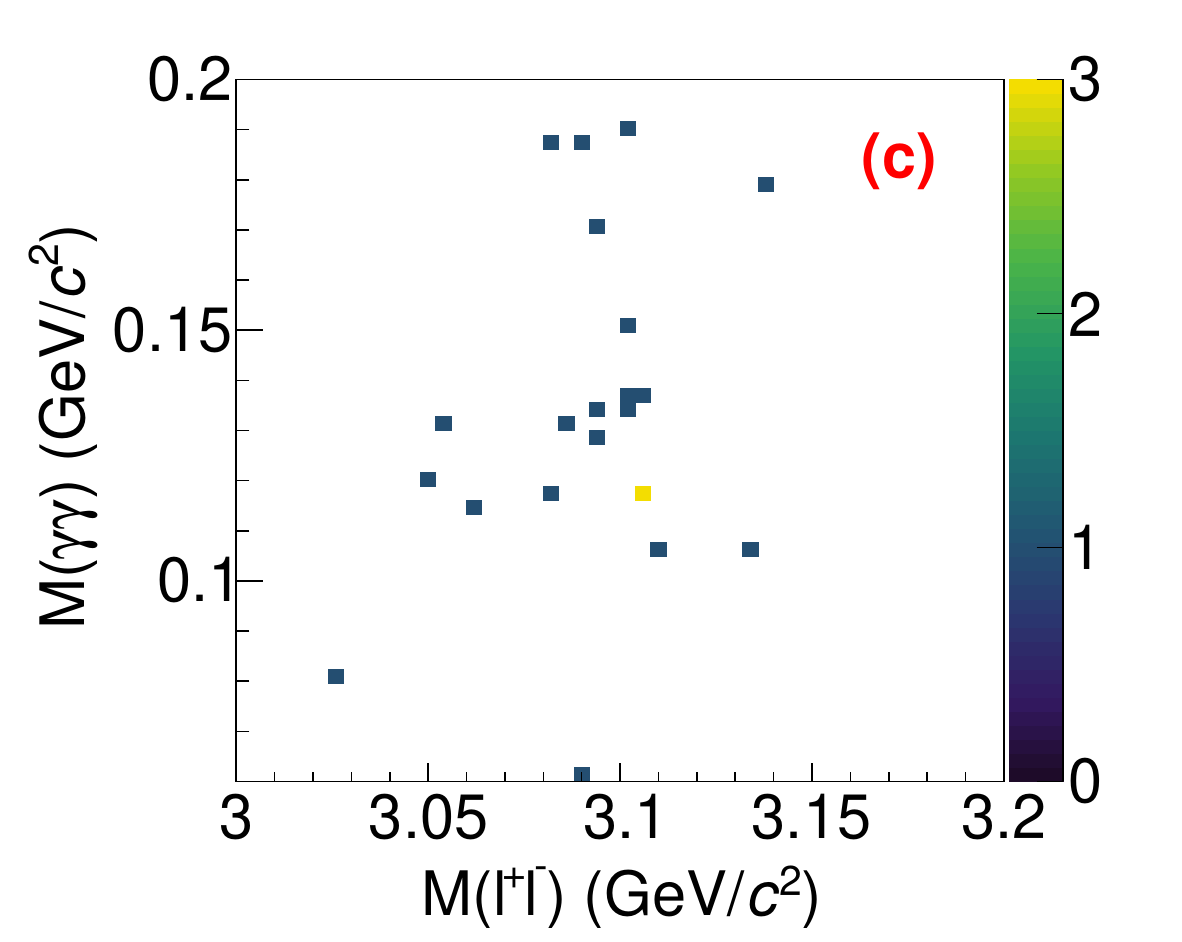}
	\includegraphics[angle=0,width=0.42\textwidth]{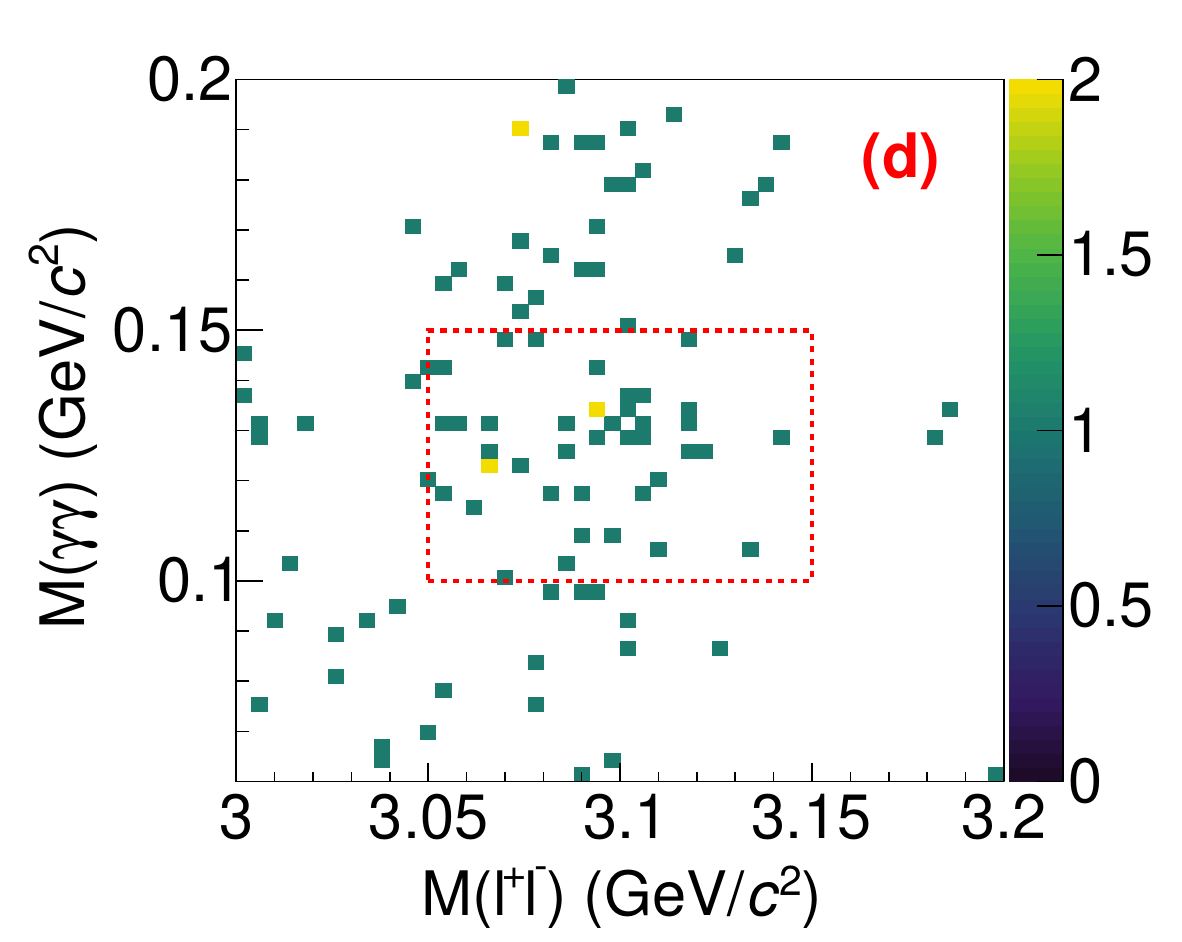}
	\caption{Invariant mass distributions of $\gamma\gamma$ versus $l^+l^-$ for signal MC (a), inclusive MC (b) and data samples (c) at the c.m.~energy of 4.682 GeV, as well as for the sum of all data samples (d). The signal region corresponds to the area inside the red dashed box, while the sideband region corresponds to the area outside this box.}
	\label{Fig:2D}
\end{figure*}

After applying all selection criteria, the remaining background is dominated by various hadronic processes. The expected background yield is estimated from inclusive MC samples. Figure~\ref{Fig:2D} shows the two-dimensional distributions of the invariant mass of photon pairs $M(\gamma\gamma)$ versus the invariant mass of lepton pairs $M(l^+l^-)$ for the signal MC, inclusive MC, and data samples at the c.m.~energy of 4.682~GeV. Since the signal and inclusive MC distributions exhibit similar features at all energy points, those at 4.682~GeV are presented as representative examples. The corresponding distribution for data, summed over all energy points, is also presented in Fig.~\ref{Fig:2D}.

\section{Cross sections and resonant structures}
\label{sec:cross_section}
\begin{figure}[!htbp]
	\centering
	\includegraphics[angle=0,width=0.45\textwidth]{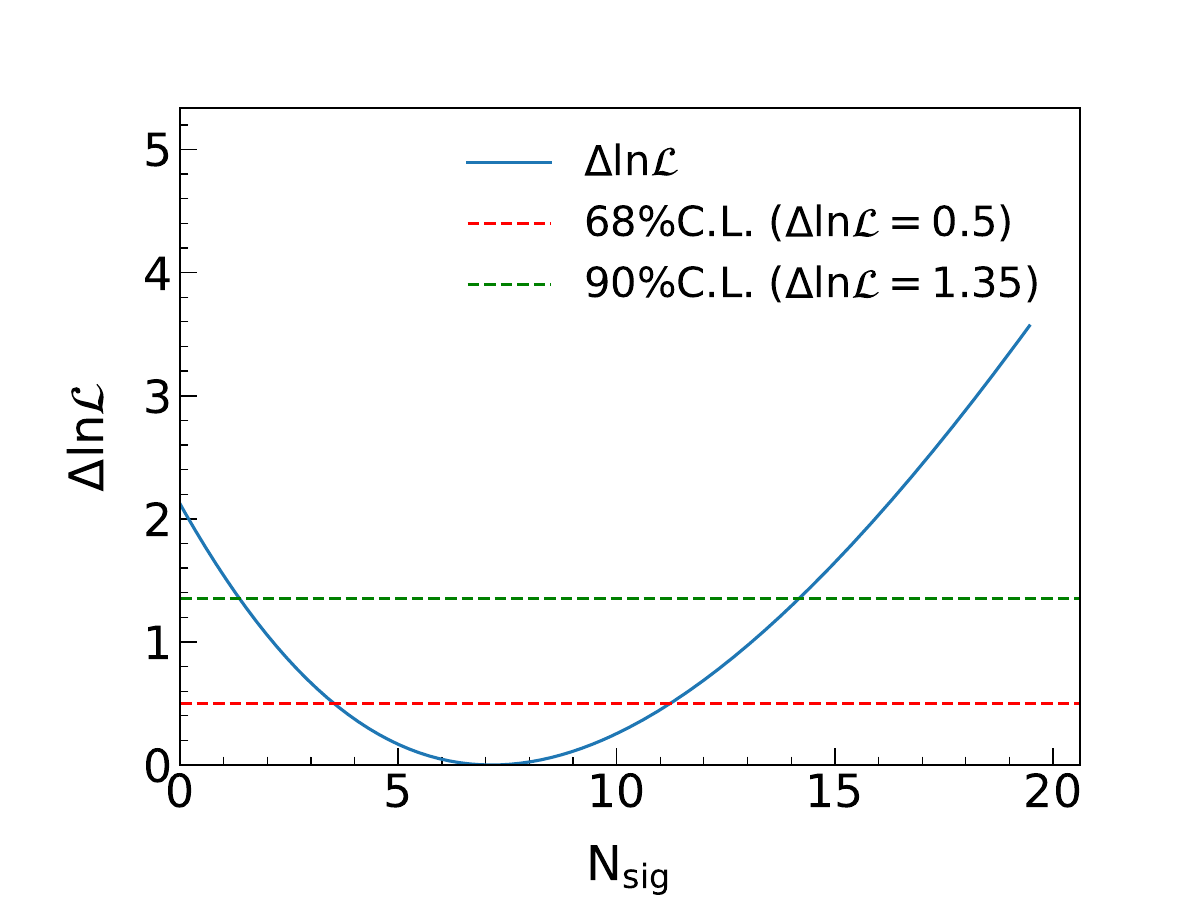}
	\caption{Likelihood scan at the c.m. energy of 4.682 GeV. The solid blue curve shows $\Delta \ln\mathcal{L}$ as a function of the signal yield, $N_{\rm sig}$, with $\Delta \ln\mathcal{L} = 0$ corresponding to the best-fit value. The red dashed line indicates the 68\% C.L., corresponding to the estimated statistical uncertainty on the signal yield, while the green dashed line indicates the 90\% C.L., corresponding to the upper limit on the signal yield.}	
	\label{Fig:lik}
\end{figure}

The $J/\psi$ signal region is defined as $M(l^+l^-) \in [3.05, 3.15]~\text{GeV}/c^2$, corresponding to approximately $\pm3\sigma$ of the $l^+l^-$ invariant mass resolution as determined from signal MC simulations. The $\pi^0$ signal region is defined as $M(\gamma\gamma) \in [0.10, 0.15]~\text{GeV}/c^2$ based on the $\gamma\gamma$ mass resolution. The sideband region is defined as $M(l^+l^-) \in [3.0, 3.2]~\text{GeV}/c^2$ and $M(\gamma\gamma) \in [0.06, 0.2]~\text{GeV}/c^2$, excluding the signal area. The yield of signal events for $e^{+}e^{-} \to \pi^0 K^+K^-J/\psi$ is determined by maximizing the likelihood function, which is constructed under the assumption that the observed event counts in the signal and sideband regions follow Poisson distributions. The likelihood function is defined as
\begin{equation}
\mathcal{L}(x, y \mid s, b, \tau) = \mathcal{P}(x \mid s + \tau b) \, \mathcal{P}(y \mid b),
\end{equation}
where $x$ and $y$ are the observed numbers of events in the signal and sideband regions, respectively; $s$ and $b$ are the expected numbers of signal and background events; $\tau$ is the ratio of the number of background events in the signal region to that in the sideband region, estimated from inclusive MC samples by considering the kinematic properties of the background processes; $\mathcal{P}$ represents the Poisson probability mass function. The uncertainty on the signal yield is obtained by determining its 68\% confidence interval, and the 90\% confidence level (C.L.) upper limit is also calculated. The statistical significance of the signal is evaluated from the difference in $-2\ln\mathcal{L}$ between the null-signal hypothesis and the best-fit signal hypothesis. The significances for most energy points are found to be less than $2\sigma$, and a local significance of $2.1\sigma$ is found at $\sqrt{s} = 4.682$~GeV (Fig~\ref{Fig:lik}). For the combined full data sample, the significance is $2.0\sigma$.

The Born cross section $\sigma^{\rm B}$ for $e^{+}e^{-} \to \pi^0 K^+K^-J/\psi$ at each c.m.~energy $\sqrt{s}$ is calculated using
\begin{equation} \label{eq:sigma}
\sigma^{\rm B} = \frac{\sigma^{\rm D}}{|1-\Pi|^{-2}} = \frac{N_{\rm sig}}{\mathcal{L}_{\rm int} \cdot \varepsilon \cdot (1+\delta)_{\rm ISR} \cdot |1-\Pi|^{-2} \cdot \mathcal{B}}.
\end{equation}
Here, $\sigma^{\rm D}$ is the dressed cross section, $N_{\rm sig}$ is the signal yield, $\mathcal{L}_{\rm int}$ is the integrated luminosity, $\varepsilon$ is the detection efficiency, $\mathcal{B} = \mathcal{B}(J/\psi \to l^+l^-) \cdot \mathcal{B}(\pi^0 \to \gamma\gamma)$ is the product of the branching fractions taken from the PDG~\cite{ParticleDataGroup:2026mpi}, $|1-\Pi|^{-2}$ is the vacuum polarization factor calculated using QED~\cite{WorkingGrouponRadiativeCorrections:2010bjp}, and $(1+\delta)_{\rm ISR}$ is the initial state radiation correction factor, determined iteratively to account for the cross section line shape~\cite{Sun:2020ehv}. The obtained Born cross sections, their upper limits, and all related quantities for each energy point are listed in Table~\ref{Tab:par}.

\begin{table*}[!htb]
\centering
\caption{Born cross sections for $e^+e^- \to \pi^0 K^+K^- J/\psi$ and related parameters: c.m.~energy $\sqrt{s}$, integrated luminosity $\mathcal{L}_{\rm int}$, detection efficiency $\varepsilon$, number of events in the signal (``sig'') and sideband (``sid'') regions, background scaling factor $\tau$, signal yield $N_{\rm sig}$, 90\% C.L. upper limit on the signal yield $N_{\rm up}$, product of branching fractions $\mathcal{B}$, ISR correction factor $(1+\delta)_{\rm ISR}$, vacuum polarization factor $|1-\Pi|^{-2}$, Born cross section $\sigma^{\rm B}$, and the corresponding 90\% C.L. upper limit $\sigma^{\rm up}$. The uncertainties in $\tau$ and $N_{\rm sig}$ are statistical, arising from the finite sizes of the inclusive MC and data samples, respectively. For the $\sigma^{\rm B}$, the first uncertainty is statistical and the second is systematic.}
\label{Tab:par}
\begin{tabular}{ccccccccccccc}
\hline\hline
$\sqrt{s}$ & $\mathcal{L}_{\rm int}$ & $\varepsilon$ & sig & sid & $\tau$ & $N_{\rm sig}$ & $N_{\rm up}$ & $\mathcal{B}$ & $(1+\delta)_{\rm ISR}$ & $|1-\Pi|^{-2}$ & $\sigma^{\rm B}$ & $\sigma^{\rm up}$ \\
(MeV) & (pb$^{-1}$) & & & & & & & & & & (pb) & (pb) \\
\hline
4260&828.4&0.02&0&3&0.35$\pm$0.05&$-1.1^{+1.2}_{-0.8}$&1.0&0.117&0.67&1.05&$-0.69^{+0.77}_{-0.56}\pm0.04$&0.66~\\
4270&531.1&0.04&0&1&0.35$\pm$0.04&$-0.3^{+1.1}_{-0.8}$&1.2&0.117&0.68&1.05&$-0.19^{+0.62}_{-0.42}\pm0.01$&0.66~\\
4280&175.7&0.06&0&0&0.35$\pm$0.04&$0.0^{+0.5}_{-0.5}$&1.4&0.117&0.69&1.05&$0.00^{+0.55}_{-0.55}\pm0.00$&1.46~\\
4290&502.4&0.08&0&0&0.35$\pm$0.05&$0.0^{+0.5}_{-0.5}$&1.4&0.117&0.71&1.05&$0.00^{+0.15}_{-0.15}\pm0.00$&0.39~\\
4315&501.2&0.10&0&1&0.35$\pm$0.04&$-0.3^{+1.1}_{-0.8}$&1.2&0.117&0.71&1.05&$-0.08^{+0.25}_{-0.17}\pm0.00$&0.25~\\
4340&505.0&0.12&0&0&0.36$\pm$0.04&$0.0^{+0.5}_{-0.5}$&1.4&0.117&0.73&1.05&$0.00^{+0.09}_{-0.09}\pm0.00$&0.23~\\
4360&543.9&0.13&0&0&0.36$\pm$0.04&$0.0^{+0.5}_{-0.5}$&1.4&0.117&0.74&1.05&$0.00^{+0.08}_{-0.08}\pm0.00$&0.20~\\
4380&522.7&0.14&0&2&0.37$\pm$0.03&$-0.7^{+1.1}_{-0.8}$&1.1&0.117&0.73&1.05&$-0.12^{+0.18}_{-0.13}\pm0.01$&0.16~\\
4400&507.8&0.14&0&1&0.37$\pm$0.03&$-0.4^{+1.1}_{-0.8}$&1.2&0.117&0.76&1.05&$-0.06^{+0.17}_{-0.12}\pm0.00$&0.16~\\
4420&1043.9&0.14&3&3&0.38$\pm$0.03&$1.9^{+2.2}_{-1.6}$&5.8&0.117&0.76&1.05&$0.14^{+0.16}_{-0.12}\pm0.01$&0.40~\\
4440&569.9&0.15&1&0&0.39$\pm$0.03&$1.0^{+1.4}_{-0.7}$&3.6&0.117&0.77&1.05&$0.12^{+0.17}_{-0.09}\pm0.01$&0.43~\\
4470&111.1&0.15&0&0&0.40$\pm$0.03&$0.0^{+0.5}_{-0.5}$&1.4&0.117&0.76&1.05&$0.00^{+0.33}_{-0.33}\pm0.00$&0.84~\\
4530&112.1&0.15&1&1&0.43$\pm$0.02&$0.6^{+1.4}_{-0.9}$&3.3&0.117&0.78&1.05&$0.35^{+0.84}_{-0.58}\pm0.02$&1.90~\\
4575&48.9&0.15&0&1&0.46$\pm$0.02&$-0.5^{+1.1}_{-0.8}$&1.2&0.117&0.79&1.05&$-0.65^{+1.58}_{-1.13}\pm0.03$&1.54~\\
4600&586.9&0.14&2&3&0.47$\pm$0.01&$0.6^{+1.9}_{-1.5}$&4.1&0.117&0.80&1.05&$0.07^{+0.23}_{-0.19}\pm0.00$&0.47~\\
4612&103.7&0.15&0&1&0.48$\pm$0.01&$-0.5^{+1.1}_{-0.8}$&1.2&0.117&0.81&1.05&$-0.30^{+0.71}_{-0.51}\pm0.01$&0.68~\\
4628&521.5&0.17&3&3&0.49$\pm$0.01&$1.5^{+2.2}_{-1.8}$&5.5&0.117&0.81&1.05&$0.18^{+0.26}_{-0.21}\pm0.01$&0.60~\\
4641&551.7&0.16&4&5&0.51$\pm$0.01&$1.5^{+2.5}_{-2.1}$&6.0&0.117&0.81&1.05&$0.17^{+0.30}_{-0.25}\pm0.01$&0.66~\\
4661&529.4&0.16&2&4&0.52$\pm$0.01&$-0.1^{+2.0}_{-1.7}$&3.5&0.117&0.81&1.05&$-0.01^{+0.24}_{-0.21}\pm0.00$&0.40~\\
4682&1667.4&0.16&12&9&0.54$\pm$0.01&$7.1^{+4.1}_{-3.6}$&14.3&0.117&0.82&1.05&$0.27^{+0.16}_{-0.14}\pm0.02$&0.51~\\
4699&535.5&0.16&5&6&0.56$\pm$0.01&$1.6^{+2.9}_{-2.5}$&6.7&0.117&0.82&1.05&$0.19^{+0.34}_{-0.29}\pm0.01$&0.74~\\
4740&163.9&0.15&2&2&0.60$\pm$0.01&$0.8^{+1.9}_{-1.6}$&4.3&0.117&0.81&1.05&$0.33^{+0.79}_{-0.65}\pm0.02$&1.68~\\
4750&366.6&0.16&3&3&0.61$\pm$0.01&$1.2^{+2.3}_{-1.9}$&5.3&0.117&0.83&1.05&$0.20^{+0.38}_{-0.32}\pm0.01$&0.83~\\
4781&511.5&0.16&1&3&0.64$\pm$0.01&$-0.9^{+1.6}_{-1.0}$&2.1&0.117&0.84&1.06&$-0.11^{+0.20}_{-0.12}\pm0.00$&0.24~\\
4843&525.2&0.16&1&3&0.72$\pm$0.02&$-1.1^{+1.7}_{-1.0}$&1.9&0.117&0.84&1.06&$-0.13^{+0.19}_{-0.11}\pm0.01$&0.21~\\
4919&207.8&0.17&1&0&0.82$\pm$0.03&$1.0^{+1.4}_{-0.7}$&3.6&0.117&0.85&1.06&$0.28^{+0.37}_{-0.19}\pm0.01$&0.96~\\
4951&159.3&0.17&1&0&0.87$\pm$0.03&$1.0^{+1.4}_{-0.7}$&3.6&0.117&0.86&1.06&$0.35^{+0.48}_{-0.25}\pm0.02$&1.24~\\
\hline\hline
\end{tabular}
\end{table*}

\begin{figure*}[t!]
\centering
\includegraphics[width=0.32\textwidth]{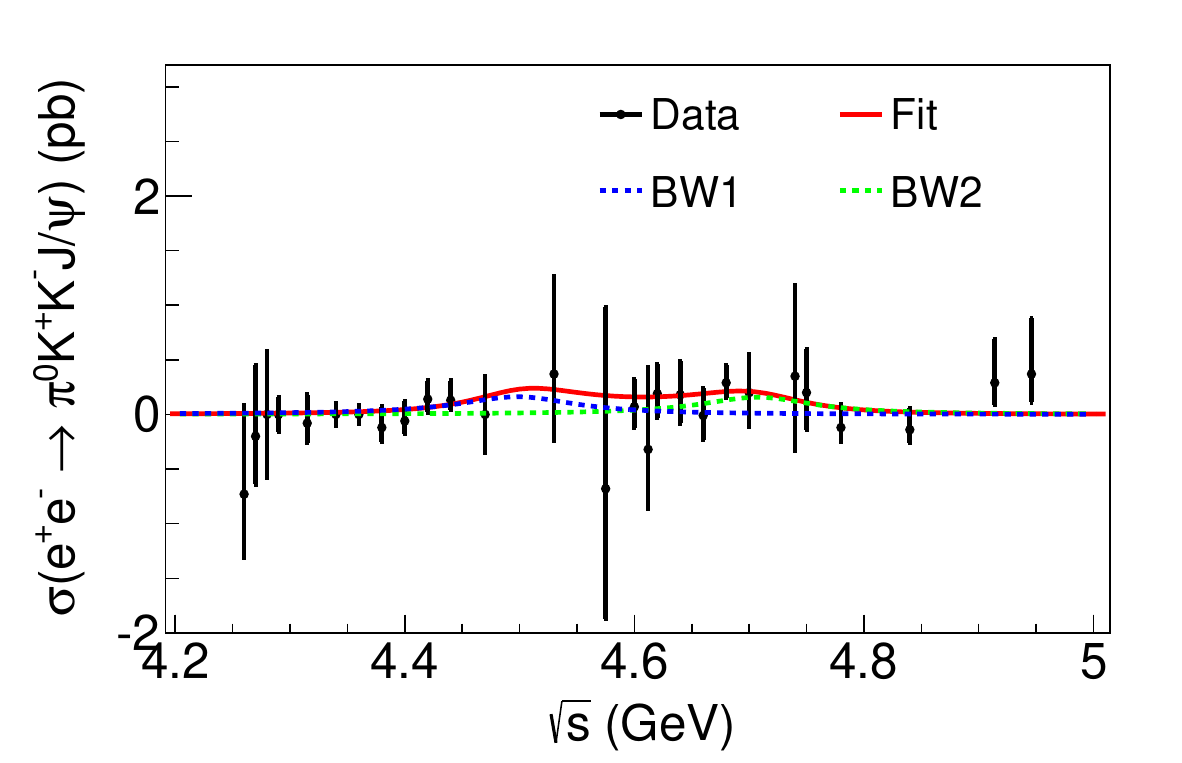}
\includegraphics[width=0.32\textwidth]{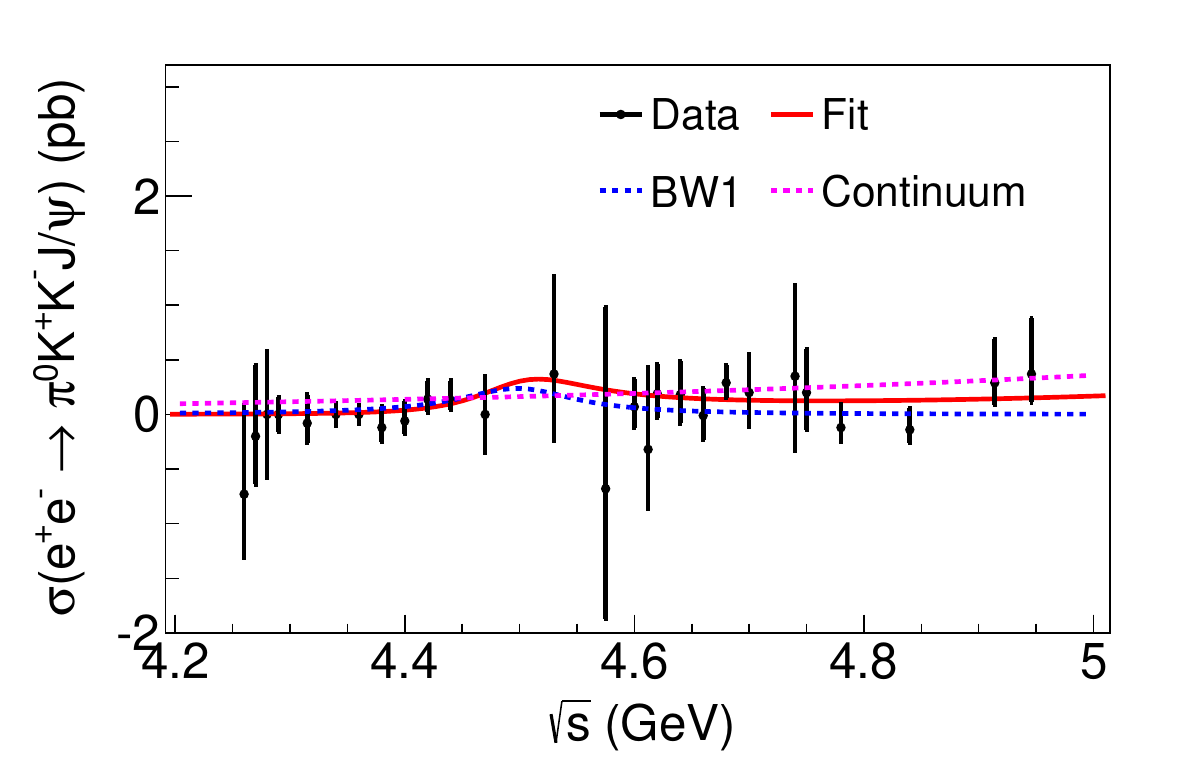}
\includegraphics[width=0.32\textwidth]{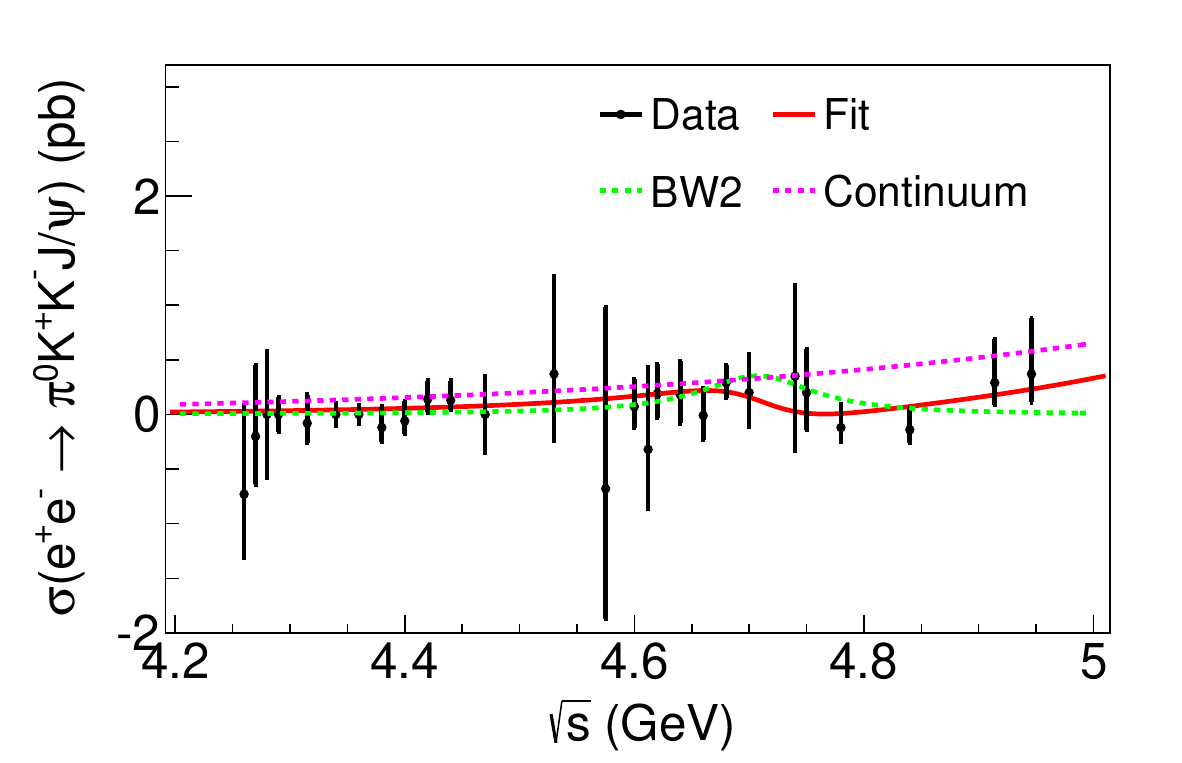}
\caption{Fits to the dressed cross section of $e^+e^- \to \pi^0 K^+K^- J/\psi$ under different hypotheses: coherent sum of the $Y(4500)$ and $Y(4710)$ BW functions (left); $Y(4500)$ plus a continuum amplitude (middle); $Y(4710)$ plus a continuum amplitude (right). The data points with error bars represent the measured cross sections with combined statistical and systematic uncertainties. The red solid curves show the total fit results. The blue and red dashed curves in the left panel show the individual contributions of the $Y(4500)$ and $Y(4710)$, respectively. The pink dashed curves in the middle and right panels show the contributions from the continuum amplitude.}
\label{Fig:isr_sys}
\end{figure*}

A maximum likelihood fit is performed on the $e^+e^- \to \pi^0 K^+K^- J/\psi$ dressed cross sections, incorporating both statistical and systematic uncertainties. The baseline model for the fit consists of a coherent sum of two relativistic Breit-Wigner (BW) amplitudes, corresponding to the hypothetical $Y(4500)$ and $Y(4710)$ resonances:
\begin{equation} \label{eq:dressed}
\sigma^{\rm D}(\sqrt{s}) = \left| \mathrm{BW}_{1}(\sqrt{s}) + \mathrm{BW}_{2}(\sqrt{s}) \, e^{i\phi} \right|^2,
\end{equation}
where
\[
\mathrm{BW}_{j}(\sqrt{s}) = \frac{M_j}{\sqrt{s}} \frac{\sqrt{12\pi \, \Gamma^{\rm tot}_{j} \, \Gamma^{ee}_{j} \, \mathcal{B}_j }}{s - M_j^2 + i M_j \Gamma^{\rm tot}_j } \sqrt{ \frac{\Phi(\sqrt{s})}{\Phi(M_j)} }.
\]
Here, $M_j$, $\Gamma^{\rm tot}_j$, $\Gamma^{ee}_j$, and $\mathcal{B}_j = \mathcal{B}(Y \to \pi^0 K^+K^- J/\psi)$ are the mass, total width, electronic partial width, and branching fraction for the $j$-th resonance ($j=1,2$), respectively. The values of mass and total width are fixed to the world average values: $(M, \Gamma^{\rm tot}) = (4484~\text{MeV}/c^2, 111~\text{MeV})$ for the $Y(4500)$~\cite{BESIII:2022joj} and $(4708~\text{MeV}/c^2, 126~\text{MeV})$ for the $Y(4710)$~\cite{BESIII:2023wqy}. The factor $\Phi(\sqrt{s})$ is the PHSP factor~\cite{ParticleDataGroup:2026mpi}, and $\phi$ is a relative phase between the two BW amplitudes in Eq.~\ref{eq:dressed}. The statistical significances of both resonant components in the fit are found to be less than $2\sigma$. The fit result is shown in the left panel of Fig.~\ref{Fig:isr_sys}.

Including an additional non-resonant continuum amplitude, parameterized as $[f/(\sqrt{s}/M)^n] \cdot [\Phi(\sqrt{s})/\Phi(M)]$, where $f$ and $n$ are free parameters and $M$ is a reference mass, does not improve the fit quality. Fits are also performed with a single resonance, either $Y(4500)$ or $Y(4710)$, interfering with the continuum amplitude described above, as shown in the middle and right panels of Fig.~\ref{Fig:isr_sys}. Under the assumption that the signal arises from a single resonance plus a continuum contribution, upper limits at the 90\% C.L. on the product $\Gamma^{ee}_{j} \mathcal{B}_{j}$ are derived by scanning the likelihood profile. Systematic uncertainties in the cross section fit are accounted for by varying the fixed resonance parameters, $M_j$ and $\Gamma^{\rm tot}_j$, within their quoted uncertainties; the largest upper limits obtained from these variations are adopted as the final results. The obtained 90\% C.L. upper limits are
\[
\Gamma^{ee} \cdot \mathcal{B} (Y(4500) \to \pi^0 K^+K^- J/\psi) < 0.038 ~\text{eV},
\]
\[
\Gamma^{ee} \cdot \mathcal{B} (Y(4710) \to \pi^0 K^+K^- J/\psi) < 0.051 ~\text{eV}.
\]

\begin{figure*}[t!]
\centering
\includegraphics[width=0.32\textwidth]{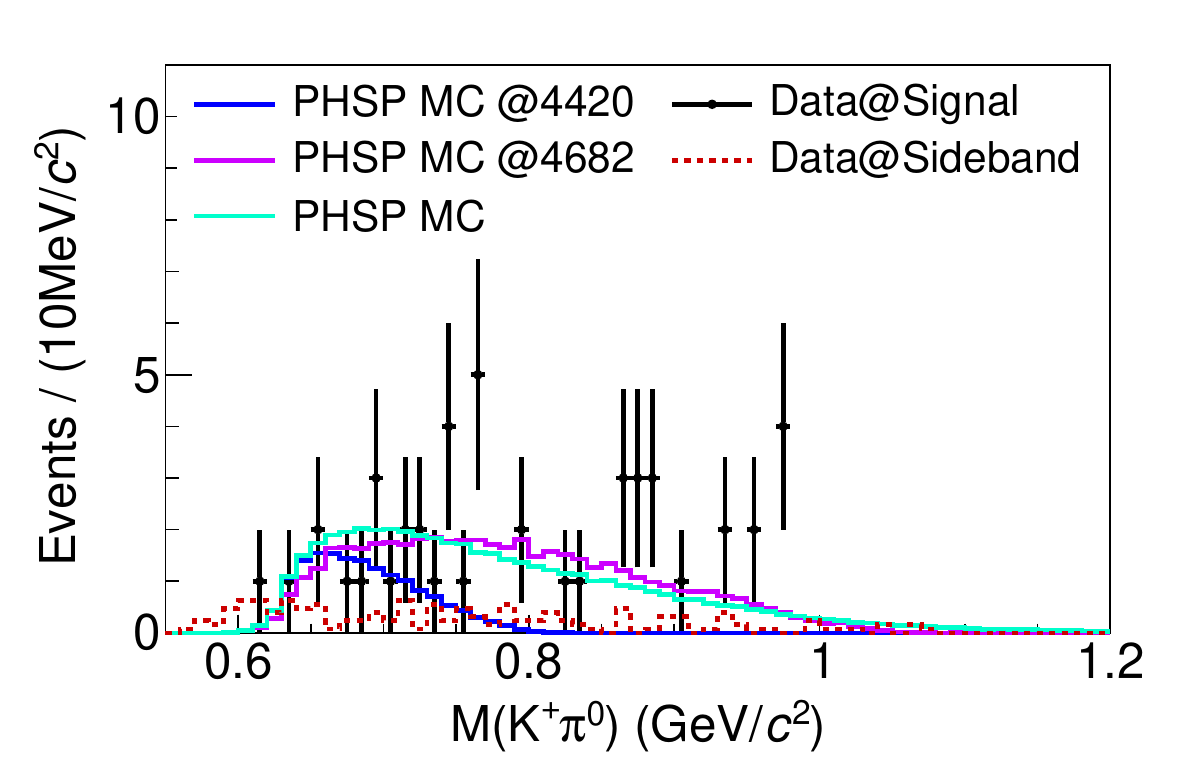}
\includegraphics[width=0.32\textwidth]{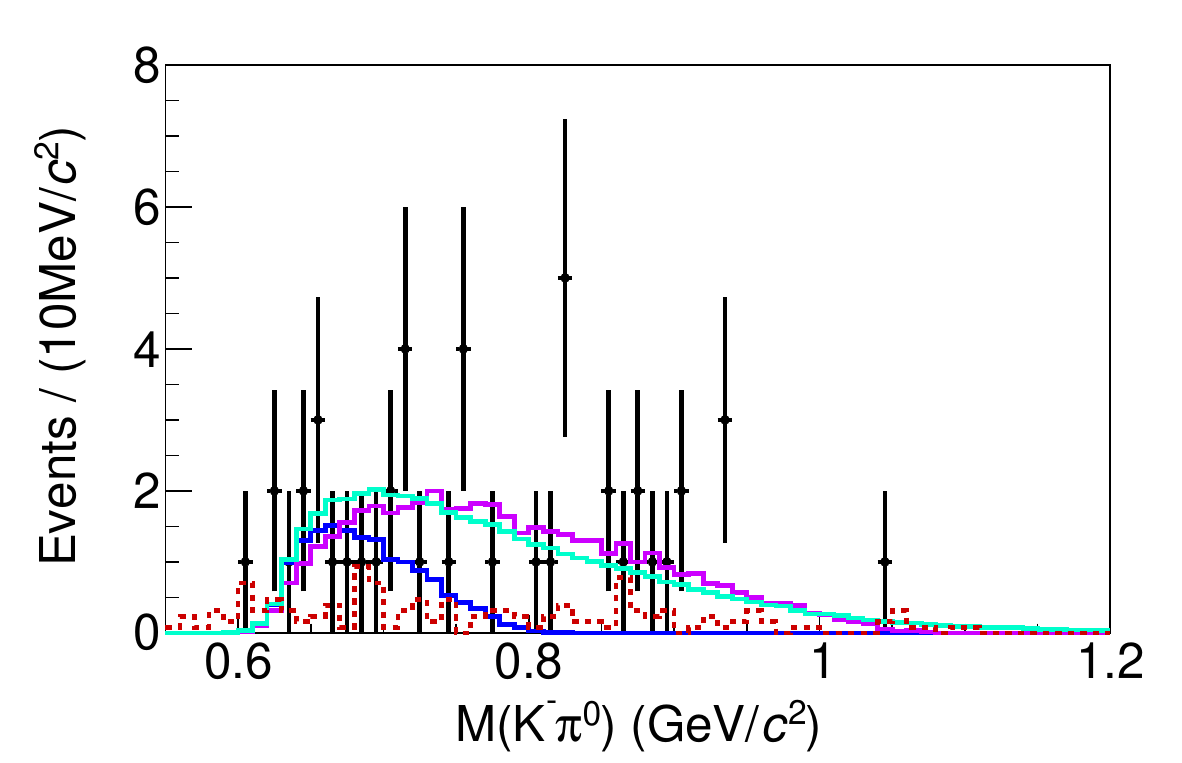}
\includegraphics[width=0.32\textwidth]{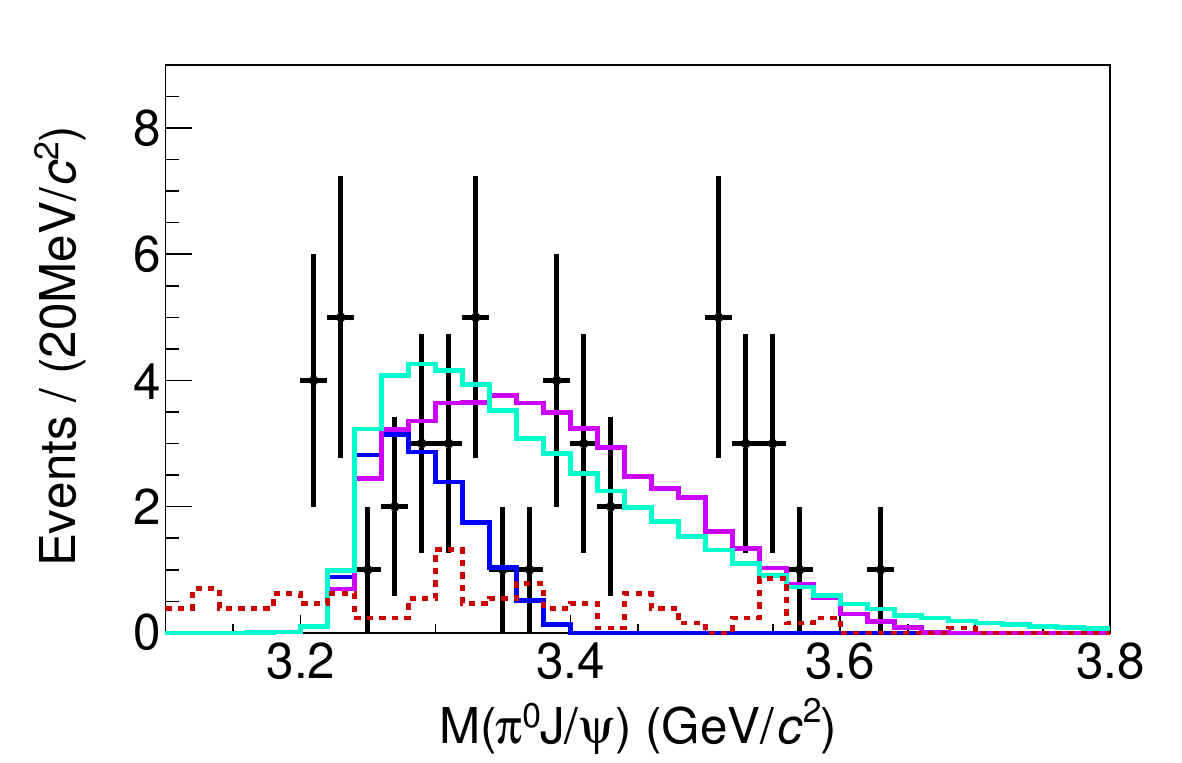}\\
\includegraphics[width=0.32\textwidth]{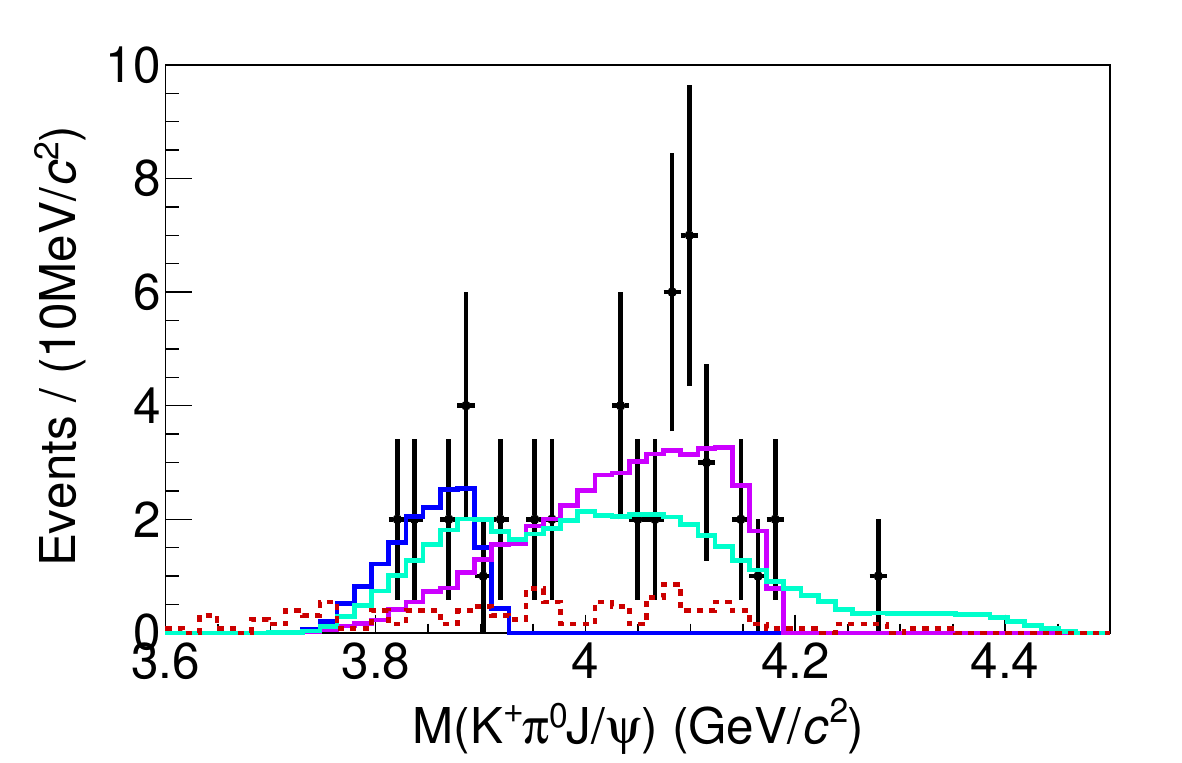}
\includegraphics[width=0.32\textwidth]{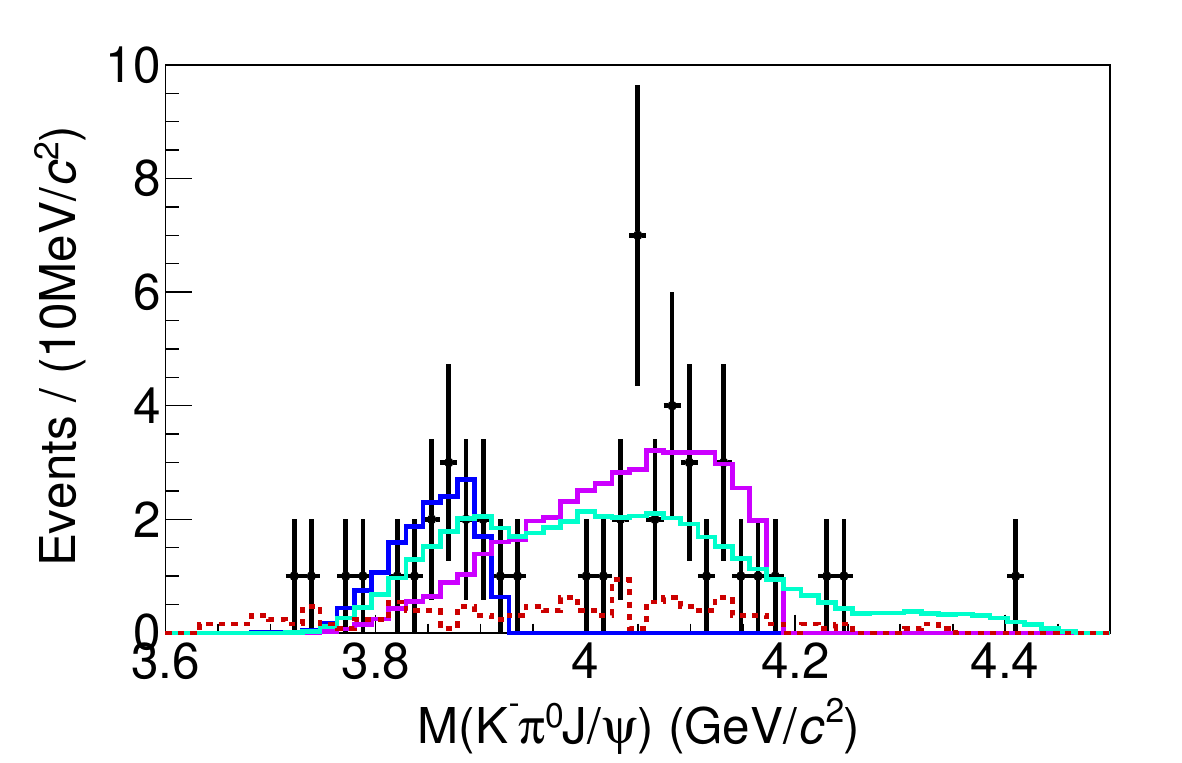}
\includegraphics[width=0.32\textwidth]{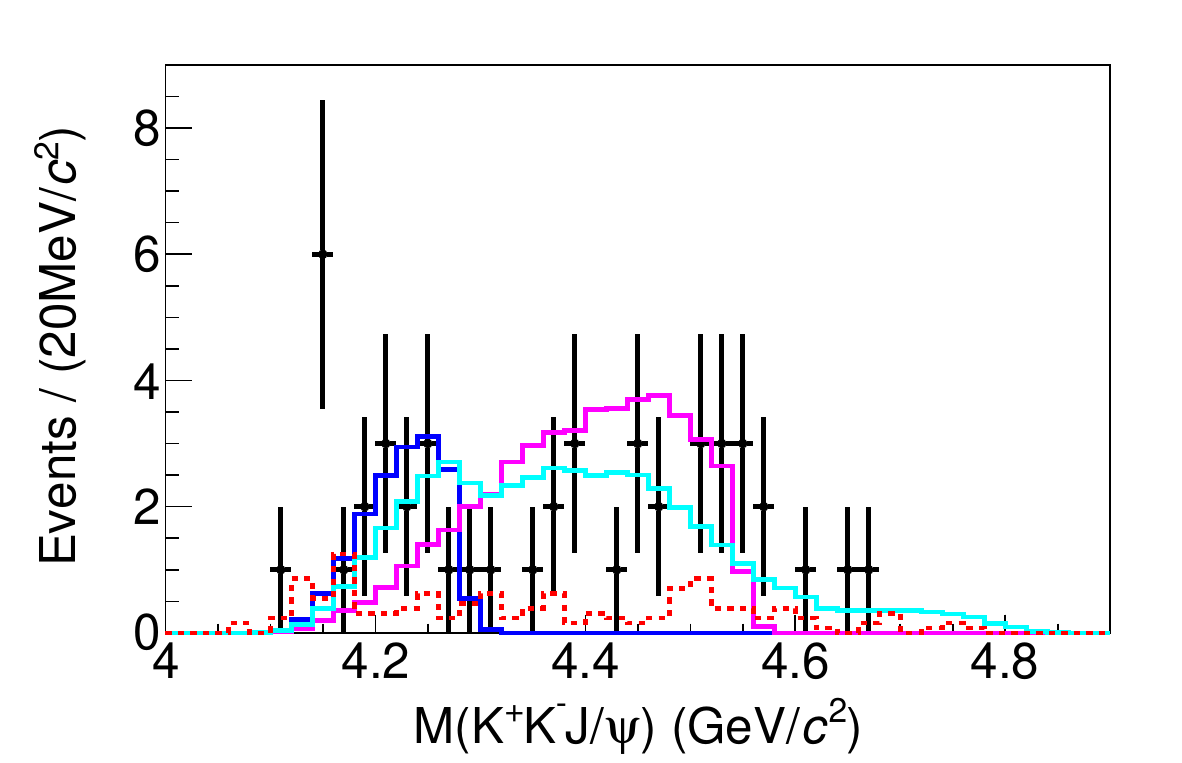}\\
\includegraphics[width=0.32\textwidth]{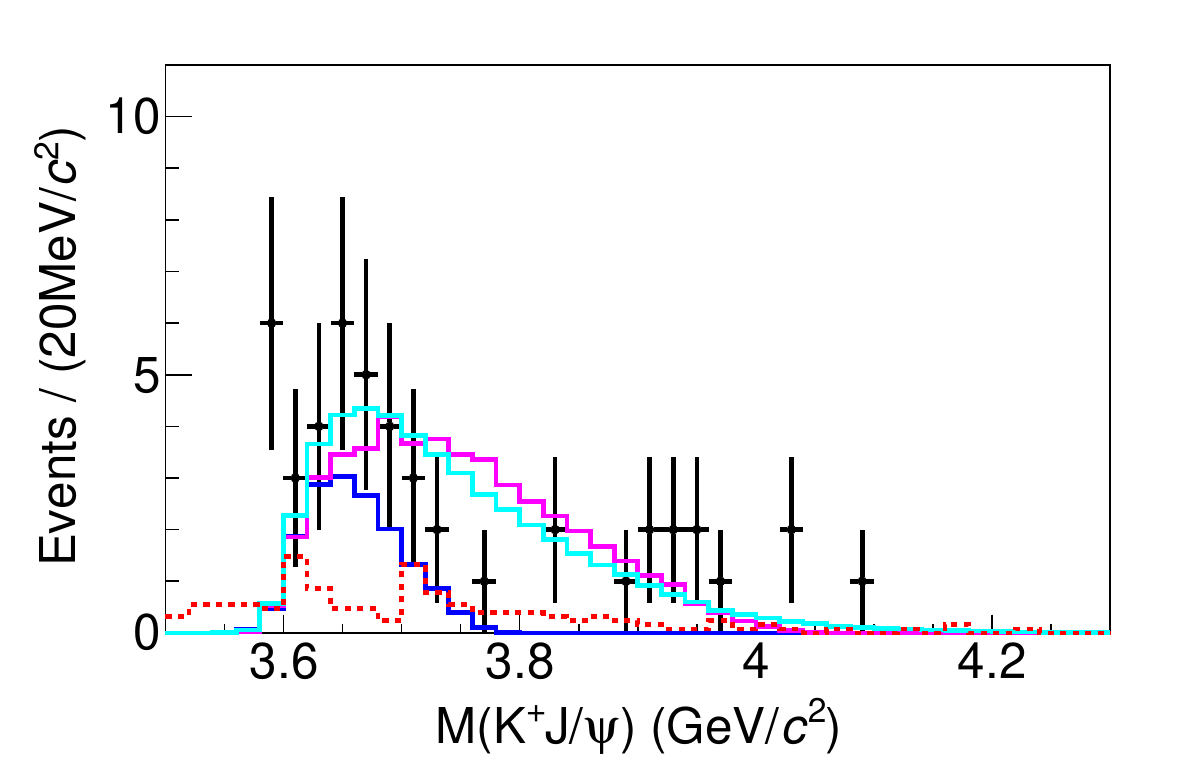}
\includegraphics[width=0.32\textwidth]{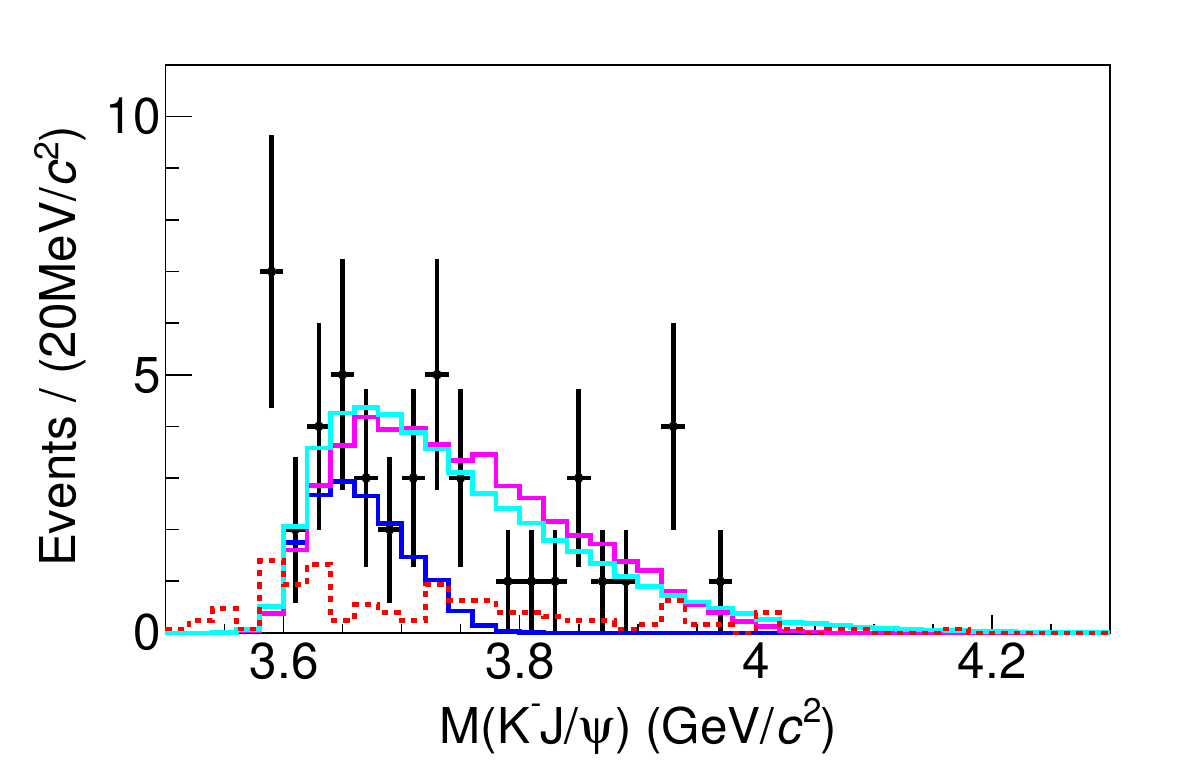}
\includegraphics[width=0.32\textwidth]{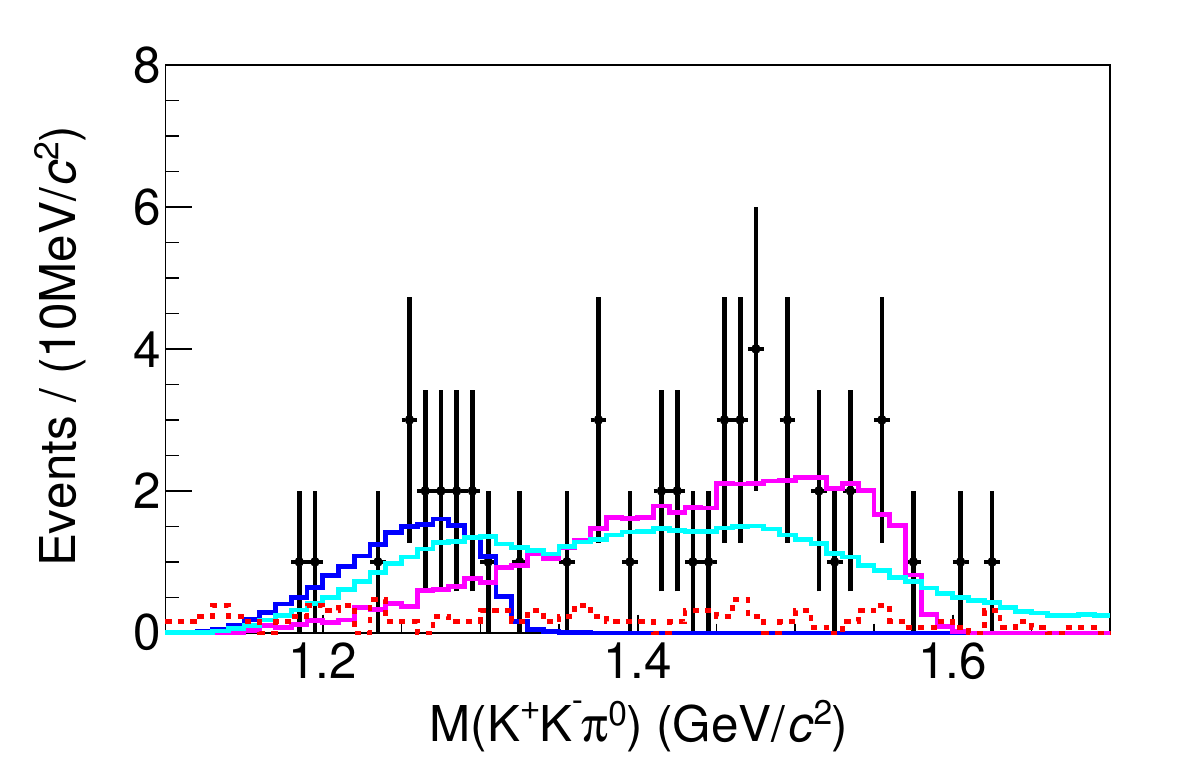}
\caption{Invariant mass distributions for the combined data sample in the signal region: $M(K^+ \pi^0)$, $M(K^- \pi^0)$, $M(\pi^0 J/\psi)$, $M(K^+ \pi^0 J/\psi)$, $M(K^- \pi^0 J/\psi)$, $M(K^+K^- J/\psi)$, $M(K^+ J/\psi)$, $M(K^- J/\psi)$, and $M_{\text{recoil}}(J/\psi)$ (i.e., $M(\pi^0 K^+ K^-)$). Data points with error bars represent the sum of all signal region data. The red dashed histograms show the sideband data. The solid cyan histograms show the PHSP MC distributions, summed over all energy points and weighted by the integrated luminosity, efficiency, and measured cross section at each point, and then normalized to the number of events in data. The solid blue and pink histograms show the PHSP MC distributions at specific c.m.~energies of 4.420 GeV and 4.682 GeV, respectively. The PHSP MC at c.m.~energies of 4.682 GeV is also normalized to the data events, while the normalization of the PHSP MC at the c.m.~energy of 4.420 GeV additionally accounts for the integrated luminosity.}
\label{Fig:state}
\end{figure*}

The invariant mass distributions of $K^+K^-J/\psi$, $\pi^0 J/\psi$, $\pi^0 K^+ K^-$, $K^\pm J/\psi$, and $K^\pm \pi^0$ are studied using the combined data samples. The recoil mass spectrum against the $J/\psi$, $M_{\text{recoil}}(J/\psi)$, is examined to search for the process $e^+e^- \to X J/\psi$, with $X \to \pi^0 K^+K^-$; the invariant mass of $K^\pm \pi^0$ $M(K^\pm \pi^0)$, is analyzed to search for contributions from $K^*$; the invariant mass of $XJ/\psi$ with $X = \pi^0, K^+K^-, K^\pm$ $M(X J/\psi)$, is inspected for $Z_{c}$ or $Z_{cs}$ signals. After comparing all distributions with the corresponding PHSP MC simulations, as shown in Fig.~\ref{Fig:state}, no significant intermediate structure is observed. Although no clear $Z_{cs}$ signal is observed the upper limits on $Z_{cs}(3985)^{\pm}$ production provide useful constraints on the properties of these states and help guide future experimental searches. The corresponding 90\% C.L.~upper limits on the production cross sections for $e^+e^- \to \pi^{0}K^{+}Z_{cs}(3985)^{-} + \text{c.c.}$ are determined using data sets collected at c.m.~energies $\sqrt{s} \ge 4.620~\text{GeV}$. The signal yield of the $Z_{cs}$ state is extracted via an unbinned maximum likelihood fit to the $M_{\rm max}(K^{\pm}J/\psi)$ distribution, where $M_{\rm max}(K^{\pm}J/\psi)$ is the larger of the invariant masses of the $K^{+}J/\psi$ and $K^{-}J/\psi$ combinations. This choice is motivated by the expected decay topology of the $Z_{cs}(3985)$, with the smaller invariant mass combination originating primarily from reflection.
In the fit, the $Z_{cs}(3985)^{\pm}$ structure is described by the MC shape of the $e^+e^- \to \pi^{0}K^{+}Z_{cs}(3985)^{-} + \text{c.c.}$ process. The background is described by a composite model comprising two components: the PHSP MC shape for the process $e^+e^- \to \pi^0 K^+K^- J/\psi$, and a shape representing all other backgrounds not originating from the $\pi^0 K^+K^- J/\psi$ final state. In the fit, the normalization of the $e^+e^- \to \pi^0 K^+K^- J/\psi$ component is allowed to float freely, while the contribution from other backgrounds is fixed to the value estimated earlier in this analysis. The 90\% confidence level upper limits on the product of the Born cross section for $e^+e^- \to \pi^{0}K^{+}Z_{cs}(3985)^{-} + \text{c.c.}$ and the branching fraction $\mathcal{B}(Z_{cs}(3985)^{\pm} \to K^{\pm}J/\psi)$, along with the corresponding quantity for each individual energy point, are summarized in Table~\ref{Tab:zcs}. All upper limits incorporate the systematic uncertainties, which will be discussed in the following section.

\begin{table*}[!htb]
	\centering							
	\setlength{\tabcolsep}{4.0pt}				
	\caption{Upper limits on the product of Born cross sections of $e^+e^- \to \pi^{0}K^{+}Z_{cs}(3985)^{-} + \text{c.c.}$ and branching fraction of $\mathcal{B}(Z_{cs}(3985)^{\pm} \to K^{\pm}J/\psi)$ ($\sigma^{\rm up} \times \mathcal{B}_{Z_{cs}}$) and related parameters: c.m.~energy $\sqrt{s}$, integrated luminosity $\mathcal{L}_{\rm ini}$, detection efficiency $\varepsilon$, 90\% C.L.~upper limit on the $Z_{cs}(3985)^{\pm}$ signal yield $N_{\rm up}$, ISR correction factor $(1+\delta)_{\rm ISR}$, and vacuum polarization factor $|1-\Pi|^{-2}$.}
	\begin{tabular}{cccccccc}
		\hline
		\hline	
		$\sqrt{s}$(MeV) & $\mathcal{L}_{\rm int}(\rm pb^{-1})$ & $\varepsilon$ &$N_{\rm up}$ &$\mathcal{B}$  & $(1+\delta)_{\rm ISR}$ & $|1-\Pi|^{-2}$ &$\sigma^{\rm up} \times \mathcal{B}_{Z_{cs}}$(pb) \\ \hline
		4628&521.5&0.19&3.77&0.117&0.68&1.05&0.46~\\
		4641&551.6&0.18&2.44&0.117&0.73&1.05&0.28~\\
		4661&529.4&0.20&2.42&0.117&0.77&1.05&0.25~\\
		4682&1667.4&0.19&2.62&0.117&0.80&1.05&0.08~\\
		4699&535.5&0.22&2.44&0.117&0.85&1.05&0.20~\\
		4740&163.9&0.14&2.42&0.117&1.49&1.05&0.57~\\
		4750&366.6&0.13&2.42&0.117&2.05&1.05&0.20~\\
		4781&511.5&0.12&4.04&0.117&1.74&1.06&0.31~\\
		4843&525.2&0.19&5.71&0.117&0.83&1.06&0.57~\\
		4918&207.8&0.22&3.48&0.117&0.80&1.06&0.75~\\
		4951&159.3&0.23&2.42&0.117&0.81&1.06&0.66~\\
		\hline\hline
	\end{tabular}
	\label{Tab:zcs}
\end{table*}

\section{Systematic uncertainties}
\label{sec:systematics}

Several sources contribute to the multiplicative systematic uncertainty in the Born cross section measurement: integrated luminosity ($\mathcal{L}_{\rm int}$), tracking, photon reconstruction, PID, the branching fractions $\mathcal{B}(J/\psi \to e^+e^-/\mu^+\mu^-)$ and $\mathcal{B}(\pi^0 \to \gamma\gamma)$, kinematic fit, ISR correction factor $(1+\delta)_{\rm ISR}$, the MUC response, and the selection criteria for the $J/\psi$ and $\pi^0$ signal regions. The individual contributions from these sources are detailed below.

The integrated luminosity is determined from Bhabha scattering events, with an associated uncertainty of 1.0\% for data collected at c.m.~energies from 4.26 to 4.60~GeV~\cite{BESIII:2022dxl} and 0.6\% for data at $\sqrt{s} > 4.60$~GeV~\cite{BESIII:2022ulv}. Considering the impact of the cross feed of full and partial reconstruction modes, the systematic uncertainty related to charged track reconstruction is estimated by reducing the tracking efficiency for each track in the MC simulation by 1.0\% and the difference in the overall signal efficiency with and without this change is taken as the uncertainty. The systematic uncertainty for photon reconstruction is assigned as 1.0\% per photon~\cite{BESIII:2010ank}, leading to a total contribution of 2.0\% for the two photons from the $\pi^0$ decay. For kaon identification, a systematic uncertainty of 1.0\% per kaon track is applied~\cite{BESIII:2013qmu}, with total values of 2.0\% and 1.0\% for the full and partial reconstruction modes, respectively. The total uncertainty of PID is calculated by combining the contributions from these reconstruction modes, weighted by their respective efficiencies. The systematic uncertainty associated with the kinematic fit is evaluated by comparing the signal efficiencies obtained from signal MC samples with and without corrections to the track helix parameters. The difference is takenas the systematic uncertainty.

The ISR correction factor $(1+\delta)_{\rm ISR}$ is derived from an assumed input cross section line shape. To estimate its systematic uncertainty, the default input line shape (e.g., a continuum model) is replaced by a model consisting of two coherent BW functions for the $Y(4710)$ and $Y(4500)$ resonances. The resulting relative difference in the product $\varepsilon \cdot (1+\delta)_{\rm ISR}$ is taken as the systematic uncertainty. The uncertainty related to muon identification is studied using a control sample of $e^+e^- \to \mu^+\mu^-$ events~\cite{BESIII:2015zbz,BESIII:2020eyu}, and the relative difference in the identification efficiency between data and simulation is assigned as the systematic uncertainty. The uncertainties on the branching fractions $\mathcal{B}(J/\psi \to l^+l^-)$ and $\mathcal{B}(\pi^0 \to \gamma\gamma)$ are taken from the PDG~\cite{ParticleDataGroup:2026mpi}.

The systematic uncertainties arising from the choice of $J/\psi$ and $\pi^0$ mass windows are evaluated using dedicated control samples. The uncertainty due to the $J/\psi$ selection is studied with $e^+e^- \to K^+K^-J/\psi$ events~\cite{BESIII:2022joj}, while the uncertainty from the $\pi^0$ selection is studied with $e^+e^- \to \pi^0\pi^0J/\psi$ events~\cite{BESIII:2020oph}. In each case, the relative change in the measured signal yield when varying the mass window defines the corresponding systematic uncertainty. 

Assuming all multiplicative systematic uncertainties are independent, they are added in quadrature to obtain the total multiplicative systematic uncertainty for the Born cross section. The contributions are summarized in Table~\ref{Tab:sys_total}. 

An additive systematic uncertainty also arises from the determination of the signal yield, stemming primarily from the uncertainty in the background scale factor $\tau$. The statistical uncertainty of the inclusive MC sample, from which $\tau$ is determined, is propagated and treated as the systematic uncertainty on $\tau$. The resulting values are listed for each energy point in Table~\ref{Tab:par}.

\begin{table}[!htbp]
	\centering
	\caption{Multiplicative systematic uncertainties (in \%) for the $e^+e^- \to \pi^0 K^+K^- J/\psi$ Born cross section measurement. The listed sources are: integrated luminosity ($\mathcal{L}_{\rm int}$), tracking efficiency, photon reconstruction, PID, branching fractions, ISR correction factor ($(1+\delta)_{\rm ISR}$), kinematic fit, MUC response, $J/\psi$ signal region selection (SR$_{J/\psi}$), and $\pi^0$ signal region selection (SR$_{\pi^0}$).}
	\label{Tab:sys_total}
	\begin{tabular}{lc}
		\hline\hline
		Source & Uncertainty (\%) \\
		\hline
		$\mathcal{L}_{\rm int}$ & 0.6 \\
		Tracking & 2.2 \\
		Photon reconstruction & 2.0 \\
		PID & 1.5 \\
		Branching fractions & 0.4 \\
		$(1+\delta)_{\rm ISR}$ & 4.2 \\
		Kinematic fit & 1.7 \\
		MUC response & 1.0 \\
		SR$_{J/\psi}$ & 0.7 \\
		SR$_{\pi^0}$ & 0.3 \\
		\hline
		Total & 5.8 \\
		\hline\hline
	\end{tabular}
\end{table}

\section{Conclusion and discussion}
\label{sec:conclusion}

We report the first study of the process $e^+e^- \to \pi^0 K^+K^- J/\psi$, using data samples collected with the \mbox{BESIII} detector at c.m.~energies from 4.26 to 4.95 GeV, corresponding to an integrated luminosity of 13.0~fb$^{-1}$. No statistically significant signal was observed at any energy point. Therefore, upper limits on the Born cross section at the 90\% C.L. are presented, along with the best-fit values. A local significance of $2.1\sigma$ is found at $\sqrt{s} = 4.682$~GeV. Under the assumption that the signal and background scale proportionally with the integrated luminosity, a fivefold increase in data statistics would be required to reach a $5\sigma$ observation.

\begin{figure}[!t]
	\centering
	\includegraphics[angle=0,width=0.45\textwidth]{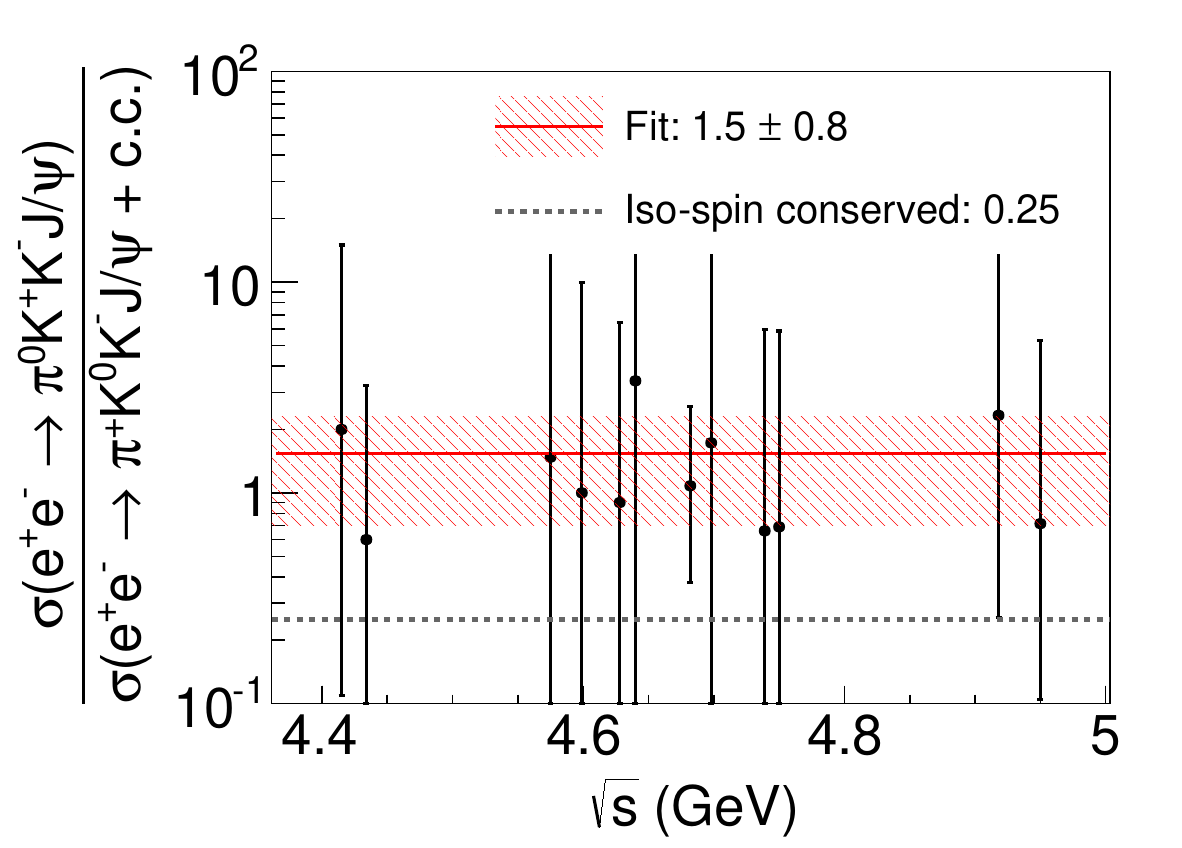}
	\caption{Ratios of the cross sections for $e^+e^- \to \pi^0 K^+K^- J/\psi$ and $e^+e^- \to \pi^{+}K^{0}K^{-}J/\psi + \text{c.c.}$ The grey dotted line indicates the expected value with isospin conservation, and the red solid line with the error band represents the constant fit result.}	
	\label{Fig:ratio}
\end{figure}

Assuming that the initial $e^+e^-$ state has zero isospin, the intermediate state effects are negligible and isospin is conserved, the theoretically expected ratio of the cross sections for $e^+e^- \to \pi^0 K^+K^- J/\psi$ and $e^+e^- \to \pi^{+}K^{0}K^{-}J/\psi + \text{c.c.}$~is 1:4. This expectation is broadly consistent with the results of the present analysis and those previously reported by BESIII~\cite{BESIII:2025fvc} (see Fig.~\ref{Fig:ratio}), although the limited statistical precision of the current measurement leads to substantial uncertainties.

No resonant structure is evident in the $\sqrt{s}$-dependence of the $e^+e^- \to \pi^0 K^+K^- J/\psi$ cross section. Under specific resonance hypotheses, upper limits at the 90\% C.L. on the products of the electronic partial width and the branching fraction are determined to be $\Gamma^{ee}_{Y(4500)}\mathcal{B}_{Y(4500)} < 0.038$~eV and $\Gamma^{ee}_{Y(4710)}\mathcal{B}_{Y(4710)} < 0.051$~eV. The ratios of these upper limits to the least reported $\Gamma^{ee}\mathcal{B}$ values for the $Y(4500)$ and $Y(4710)$ resonances from the $e^+e^-\to K^+K^-J/\psi$ channel~\cite{BESIII:2022kcv,BESIII:2023wqy} are approximately 1:10 and 1:3, respectively. Furthermore, no significant evidence is found for intermediate $Z_{c}$ or $Z_{cs}$ states in the relevant invariant mass distributions. Upper limits on the product of the Born cross sections for $e^+e^- \to \pi^{0}K^{+}Z_{cs}(3985)^{-} + \text{c.c.}$ and the branching fraction $\mathcal{B}(Z_{cs}(3985)^{\pm} \to K^{\pm}J/\psi)$ are determined at the 90\% C.L.

The ongoing upgrade of the BEPCII collider~\cite{BESIII:2020nme} is expected to deliver larger data samples in the relevant energy region. The resulting increase in statistical precision will enable more definitive studies of this process and allow for stronger constraints on the properties of exotic charmonium-like states.

\section*{Acknowledgement}
The BESIII Collaboration thanks the staff of BEPCII (https://cstr.cn/31109.02.BEPC) and the IHEP computing center for their strong support. This work is supported in part by National Key R\&D Program of China under Contracts Nos. 2025YFA1613900, 2023YFA1606000, 2023YFA1606704; National Natural Science Foundation of China (NSFC) under Contracts Nos. 12375083, 11635010, 11935015, 11935016, 11935018, 12025502, 12035009, 12035013, 12061131003, 12192260, 12192261, 12192262, 12192263, 12192264, 12192265, 12221005, 12225509, 12235017, 12342502, 12361141819, 12535005; the Chinese Academy of Sciences (CAS) Large-Scale Scientific Facility Program; the Strategic Priority Research Program of Chinese Academy of Sciences under Contract No. XDA0480600; CAS under Contract No. YSBR-101; 100 Talents Program of CAS; The Institute of Nuclear and Particle Physics (INPAC) and Shanghai Key Laboratory for Particle Physics and Cosmology; Agencia Nacional de Investigaci\'on y Desarrollo de Chile (ANID), Chile under Contract No. ANID CCTVal CIA250027; Istituto Nazionale di Fisica Nucleare, Italy; Knut and Alice Wallenberg Foundation under Contracts Nos. 2021.0174, 2021.0299, 2023.0315; Ministry of Development of Turkey under Contract No. DPT2006K-120470; National Research Foundation of Korea under Contract No. RS-2026-25486791; National Science and Technology fund of Mongolia; Polish National Science Centre under Contract No. 2024/53/B/ST2/00975; STFC (United Kingdom); Swedish Research Council under Contract No. 2019.04595; U. S. Department of Energy under Contract No. DE-FG02-05ER41374

\end{document}